\documentclass[aps,pra,twocolumn,superscriptaddress,showpacs]{revtex4-1}

\usepackage{graphics}
\usepackage{graphicx}
\usepackage{dcolumn}
\usepackage{bm}
\usepackage{color}
\usepackage{cancel}
\usepackage{amssymb,amsmath}

\begin{document}

\title{Soliton dynamics in gas-filled hollow-core photonic crystal fibers}

\author{Mohammed F. Saleh}
\affiliation{School of Engineering and Physical Sciences, Heriot-Watt University, EH14 4AS Edinburgh, UK}
\affiliation{Department of Mathematics and Physics Engineering, Alexandria University, Alexandria, Egypt}

\author{Fabio Biancalana}
\affiliation{School of Engineering and Physical Sciences, Heriot-Watt University, EH14 4AS Edinburgh, UK}

\begin{abstract}
Gas-filled hollow-core photonic crystal fibers offer unprecedented opportunities to observe novel nonlinear phenomena. The various properties of gases that can be used to fill these fibers give additional degrees of freedom for investigating nonlinear pulse propagation in a wide range of different media. In this review, we will consider some of the the new nonlinear interactions that have been discovered in recent years, in particular those which are based on soliton dynamics.
\end{abstract}

\maketitle

\section{Introduction}

A soliton is a nonlinear localized wave possessing a particle-like nature, which maintains its shape during propagation, even after an elastic collision with another soliton. In optics, this special wave-packet can arise due to the balance between nonlinear and dispersive effects. Based on confinement in time or space domain, one can have either temporal or spatial solitons. Both kinds of optical solitons can occur due to the third-order nonlinearity, named as Kerr effect \cite{Weinberger08} that leads to an intensity-dependent refractive index of the medium. This nonlinear dependence results in spatial self-focusing and temporal self-phase modulation \cite{Boyd07}. A spatial soliton is formed when the self-focusing effects counteracts the natural diffraction-induced broadening of the pulse. Similarly, a temporal soliton is developed when the self-phase modulation effects compensates the usual dispersion-induced broadening.
 
The possibility of soliton propagation in the anomalous dispersion regime of an optical fiber was predicted by analyzing theoretically the nonlinear Schr\"{o}dinger equation (NLSE) \cite{Hasegawa73}. Due to the lack of ultrashort pulses in this regime, the experimental verification was delayed until 1980, when Mollenauer \textit{et al.} were able not only to excite a fundamental soliton but also a higher-order soliton \cite{Mollenauer80}. A \textit{N}th-order soliton that represents a joint state of \textit{N} fundamental solitons, can propagate in a periodic way consisting of pulse splitting followed by a recovery to the original pulse after a characteristic propagation length.

\textit{Supercontinuum generation}, which is a massive pulse broadening, was first observed in step-index silica-core optical fibers by pumping in the normal dispersion regime \cite{Lin76} due to mutual interaction between the self-phase modulation effect and Raman-scattering process \cite{Shimizu67}. After the availability of ultrashort sources in the anomalous dispersion regime, experiments show that soliton emission leads to broadband supercontinuum generation in optical fibers \cite{Beuad87,Neto88,Islam89a,Islam89b,Schutz93}. In this case, an energetic pulse or a \textit{N}th-order soliton is continuously temporally compressed and spectrally broadened due to the interplay between the anomalous dispersion and Kerr effect. After a certain propagation distance, the pulse breaks up and a series of fundamental solitons are ejected in a process known as \textit{soliton fission} due to higher-order dispersion effects. These latter effects are also responsible to stimulate the transfer of energy from each fundamental soliton to a weak narrowband dispersive wave in the normal dispersion regime \cite{Wai86}. The energies of the higher-order soliton was shown to be equal to the sum of energies of its constituents \cite{Kodama87}. Because of the intrapulse Raman scattering of silica, the central frequency of the fundamental solitons are continuously downshifted during propagation \cite{Gordon86,Mitschke86}. This frequency shift continues until it saturates when the soliton gets chirped \cite{Agrawal07}.

\textit{Photonic crystal fibers} (PCFs) are microstructured fibers that can be engineered in various ways to tailor their linear and nonlinear properties.  PCF structures are fabricated using different techniques. The most widely used method is based on stacking a number of capillary tubes in a suitably shaped preform to form the desired arrangement, which is then drawn down to a fiber. This has led to the fabrication of the first silica-air endlessly single mode PCF \cite{Birks97}. Followed by a series of successful experiments, PCFs with special characteristics  have been demonstrated such as PCFs with large mode area \cite{Knight98}, dispersion controlled \cite{Mogilevtsev98,Knight00}, hollow-core \cite{Cregan99}, multicore \cite{Mangan00}, and birefringence \cite{Blanch00}. Another different technique to manufacture PCFs (so-called Omniguide fibers) is to deposit chalcogenide glass on a polymer using thermal evaporation, then wrap it to form multilayered Bragg fibers \cite{Yeh76,Johnson01,Abouraddy07}. A full thorough review of the history, fabrication, theory, numerical modeling, and applications of PCFs are found in \cite{Russell06}.

\textit{Solid-core} PCFs guides light via total internal reflection similar to step-index fibers. However, the additional degrees of freedom provided by modifying the capillary diameters open up different possibilities to engineer the fiber properties, such as shifting its zero dispersion wavelength (ZDW) \cite{Mogilevtsev98}, or  enhancing its Kerr nonlinearity via reducing its effective-core area \cite{Broderick99}. A number of simultaneous experiments have succeeded in exploiting these new advantages offered by solid-core PCFs in generating broadband supercontinuum   \cite{Knight00,Birks00,Ranka00a,Ranka00b}.  Supercontinuum generation in solid-core PCFs has been reviewed in \cite{Dudley06}

\textit{Hollow-core} (HC) PCFs have also attracted much interest since their invention \cite{Knight96,Cregan99,Russell03}, because of their potential for lossless and distortion-free transmission, particle trapping, optical sensing, and  novel applications in nonlinear optics \cite{Benabid02,Knight03,Ouzounov03}. HC-PCFs can be classified into two different categories based on the guiding mechanism. Photonic bandgap (PBG) HC-PCFs guide light using the concept of forbidden bands in periodic solid crystals. If the optical frequency lies within the photonic bandgap, propagation through the fiber cladding is prohibited and light is confined inside the hollow-core. These fibers offer very low-loss over a restricted wavelength bands \cite{Roberts05}.  As the number of cladding layers increases, light confinement becomes stronger. The other category is the anti-resonant fibers, such as the Kagome-lattice fiber, which can allow waveguiding in a thick low-index core surrounded by a thin high-index cladding via the anti-resonant Fabry-Perot  mechanism \cite{Duguay86}. The fibers have a broad transmission band in comparison to PBG fibers, however, with much higher but tolerable losses. Adding more cladding layers does not play a significant role in reducing these losses. A lot of work has been recently dedicated towards a dramatic suppression of the attenuation coefficient of the anti-resonant fibers, while maintaining its broad transmission bandwidth, by inducing fiber-cores with negative-curvatures \cite{Wang11,Yu12}. 

\textit{Gas-filled HC-PCFs with Kagome lattice} have become a powerful alternative for solid-core PCFs especially for nonlinear applications \cite{Russell14}. The  low-loss wide transmission window, the pressure-tunability dispersion, and the variety of different gases with special properties have offered several opportunities for demonstrating new nonlinear applications \cite{Travers11} such as Stokes generation with a drastical reduction in Raman threshold \cite{Benabid02a}, high harmonic generation \cite{Heckl09}, efficient deep-ultraviolet radiation \cite{Joly11}, ionization-induced soliton self-frequency blueshift \cite{Chang11,Hoelzer11b,Saleh11a,Chang13}, strong asymmetrical self-phase modulation, universal modulational instability \cite{Saleh12},  parity-time symmetry \cite{Saleh14}, and temporal crystals \cite{Saleh15a,Saleh15b}. The subject of this review is to present in detail some of these applications, especially those which involve soliton dynamics.

This review paper is organized as follows. In Sec. II, we give an overview of the governing equation of nonlinear pulse propagation in Kerr media and the effect of Raman-nonlinearity in silica-glass. Secs. III and IV are dedicated to the study of soliton dynamics in HC-PCFs filled by Raman-active and Raman-inactive gases, respectively. Conclusions and final remarks are presented in Sec. V.

\section{Nonlinear pulse propagation in guided Kerr media}
Nonlinear pulse propagation in a lossless Kerr medium can be described by the scalar nonlinear Schr\"{o}dinger equation (NLSE),
\begin{equation}
i\partial_{z}A+\sum_{m=2}\dfrac{i^{m}}{m!}\beta_{m}\partial_{t}^{m}A+\gamma|A|^{2}A   =0, \label{eqNLSE}
\end{equation}
where the slowly varying envelope approximation (SVEA) is assumed, $A\left(z,t \right) $ is the complex envelope of the electric field in units of W$ ^{1/2} $, \textit{z} is the longitudinal coordinate along the fiber, \textit{t} is time in a reference frame moving with the pulse group velocity, $ \beta_{m} $ is the $ m $th order dispersion coefficient calculated at the pulse central frequency $ \omega_{0} $, and $ \gamma $ is the nonlinear Kerr coefficient. This equation is usually integrated using the split-step Fourier method \citep{Agrawal07}. The second term is usually taken as a fit of the dispersion in  the frequency domain $ \omega $ as $ \sum_{m=2}\beta_{m}\left(\omega-\omega_{0} \right)^{m}/m!  $ or in an approximation-free manner as $ \beta\left(\omega \right)-\beta_{0}-\beta_{1}\left(\omega-\omega_{0} \right) $ in those cases when the dispersion relation $\beta(\omega)$ of the waveguide under consideration is known. In a regime of deep anomalous dispersion, i.e. for $\left|\beta_{2}\right|\gg\left|\beta_{m>2} \right| $, the normalized solution of Eq. (\ref{eqNLSE}) is the fundamental Schr\"{o}dinger soliton,
\begin{equation}
\psi\left( \xi,\tau\right) = N \mathrm{sech}\left( N\tau\right) \mathrm{exp}\left(iN^{2}\xi/2 \right),
\end{equation}
where $ N $ is an arbitrary parameter that controls the soliton amplitude and width, $ \xi=z/z_{0} $, $ \tau=t/t_{0} $, $ \psi=A/A_{0} $, $ A^{2}_{0}=1/\left( \gamma z_{0}\right)  $,   $ z_{0}= t_{0}^{2}/\left|\beta_{2} \right|$ is the second-order dispersion length at $\omega_{0}$, and $t_{0}$ is the input pulse duration.

There are some limitations in using the above NLSE, for instance: i) when studying ultrashort pulses, with few optical cycles ($\omega_{0}t_{0}\sim 1$), since the SVEA is no longer valid, and ii) investigating polarization effects in birefringent waveguides requires to take into account the vector nature of the electric field. Alternative, more accurate methods of simulating the propagation of electromagnetic pulses in dielectric media are the finite difference time domain (FDTD) method \cite{Goorjian92,Joseph93,Ziolkowski93,Goorjian97,Nakamura05} or the recent unidirectional pulse propagation equation (UPPE) \cite{Kolesik04,Kinsler10}. However, accuracy is achieved at the expense of an increased computational effort, leading to a deficiency of understanding different physical mechanisms behind the pulse dynamics. 

There are also several advantages in using the NLSE: i) suitability in generalizing the NLSE to include different phenomena such as Raman nonlinearity, self steepening, and photoionization effects; and (ii) the possibility of using well-known analytical techniques such as variational perturbation theory to study new nonlinear effects \cite{Agrawal07}. Based on our experience, the NLSE method can usually produce very good qualitative and quantitative results, usually accompanied by a deep physical understanding of pulse dynamics. In this review, it is our aim to present recent and novel SVEA equations that are able to provide a full understanding of the salient dynamics of pulse propagation in gas-filled fibers.

\paragraph*{Intrapulse Raman scattering redshift---} 
In  a molecular medium, a fraction of optical power can be transferred from one pulse to another via Raman effect, when the frequency difference between the two pulses matches with the vibrational modes of the medium. This effect can occur within a single pulse, when it is too short and has a broad spectrum that exceeds the Raman-frequency shift. In this case, the high-frequency (blue) spectral components of the pulse continues to amplify the low-frequency (red) components during propagation. This amplification appears as a redshift of the pulse spectrum, known as intrapulse Raman redshift. The NLSE can be modified to study this effect in silica-core fibers \cite{Agrawal07},
\begin{equation}
i\partial_{\xi}\psi+\sum_{m=2}\frac{i^{m}z_{0}}{t_{0}^{m}m!}\beta_{m}\partial_{\tau}^{m} \psi+|\psi|^{2}\psi-\tau_{R}\,\psi\,\partial_{\tau} |\psi|^{2}=0,\label{eqNLSER}
\end{equation}
where $ \tau_{R} $ is the Raman coefficient in normalized units. To investigate this  effect, higher-order dispersion coefficients are first neglected $ \left(\beta_{m>2}\approx 0 \right)  $ for simplicity. For weak Raman nonlinearity, and absence of higher-order dispersion coefficients $ \beta_{m>2} $, the solution of Eq. (\ref{eqNLSER}) can still be assumed to be a fundamental soliton that is perturbed by the Raman effect, i.e. $ \psi\left( \xi,\tau\right)=N\,\mathrm{sech} \left[N \left(\tau-\bar{\tau}\left(\xi \right) \right) \right] \exp\left[-i\delta\left(\xi \right)\tau\right] $, where  $   \delta $ and $\bar{\tau}  $ are the soliton central frequency and temporal peak that change during propagation because of Raman scattering.  Applying perturbation theory \cite{Agrawal07}, the soliton is found to be linearly redshifting in the frequency domain with rate $ g =\frac{8}{15}\tau_{R} N^{4}$, and decelerating in the time domain, i.e. $ \delta = - g\, \xi,$ and $ \bar{\tau} = g\, \xi^{2}/2 $. 

\section{Raman effect in gas-filled HC-PCFs}
Unlike silica glass, which has a very broad Raman spectrum, stimulated Raman scattering processes in gases have a narrow Raman-gain spectrum. For this reason, gases  are characterized by having a very long molecular coherence relaxation (dephasing) time, of the order of hundreds of picoseconds or even more, in comparison to the short relaxation time of phonon oscillations in silica glass (approx. 32 fs). Within this long relaxation stage, the medium can exhibit a highly non-instantaneous response to pulsed excitations. Raman responses can be manifested in either rotational or vibrational modes.  Nonlinear interactions between optical pulses and Raman-active gases have been usually exploited in the synthesis of subfemtosecond pulses \cite{Yoshikawa93,Kaplan94,Kawano98,Nazarkin99,Kalosha00}. 

Conventional approaches have been used to enhance Raman scattering in gases such as focusing a laser beam into a gas confined in a bore-fiber capillary \cite{Rabinowitz76}, or employing a gas-filled high-finessse Fabry-Perot cavity to increase the interaction length \cite{Meng00}. However, Benabid \textit{et al.} have exploited the benefits of HC-PCFs and have demonstrated Stokes-generation via Raman scattering using only few microjoule optical pulses in hydrogen-filled HC-PCF \cite{Benabid02}. The range of the used energies in this experiment were nearly two-order of magnitude less than other values reported in prior techniques; demonstrating the capability of HC-PCFs to enhance nonlinear Raman interactions in gases.

F. Belli \textit{et al.} have shown an ultrabroadband supercontinuum generation spanning from 125 nm in the vacuum-UV to 1200 nm in the mid infrared regime, when pumping a H$ _{2} $-filled photonic crystal fiber using  a 30 fs  pulse centered at wavelength 805 nm with energy 2.5 $ \mu $J \cite{Belli15}. The uniqueness of this work is the extension of supercontinuum generation below 200 nm.  Due to the interplay between the Kerr effect, the Raman responses of both the rotational and vibrational  excitations, and the shock effect  along the fiber with different levels of pulse compression, a dispersive wave at 182 nm on the trailing edge of the pulse is emitted that broadens the spectrum into the  vacuum-UV region.
 
\paragraph*{Density matrix theory---} The dynamics of the Raman polarization (also called {\em coherence}) $ P_{\mathrm{R}} $ due to a single Raman mode excitation in gases can be determined by solving the Bloch equations for an effective two-level system \cite{Kalosha00,Butylkin89},
\begin{equation}
\begin{array}{l}
\partial_{t} w + \dfrac{w+1}{T_{1}} =\dfrac{i\alpha_{12}}{\hbar}\left(\rho_{12}-\rho_{12}^{*} \right)E^{2}, \\ 
\left[ \partial_{t}  + \dfrac{1}{T_{2}}-i\omega_{\rm R}\right]\rho_{12} =\dfrac{i}{2\hbar}\left[\alpha_{12} w + \left(\alpha_{11}-\alpha_{22} \right)\rho_{12} \right]E^{2},
\end{array}
\end{equation}
where $ \alpha_{ij} $ and $ \rho_{ij} $ are the elements of the $ 2\times 2 $ polarizability and density matrices, respectively, $ E\left(z,t \right) $ is the real electric field, $ \omega_{\rm R} $ is the Raman frequency of the transition, $ w=\rho_{22}-\rho_{11} $ is the population inversion between the excited and ground states, $ \rho_{22}+\rho_{11}=1 $,  $ \rho_{21}=\rho_{12}^{*} $, $ \alpha_{12}=\alpha_{21}$,  $ N_{0} $ is the molecular number density, $ T_{1} $ and $ T_{2} $ are the population and polarization relaxation times, respectively, and $ \hbar $ is the reduced Planck's constant. Solving these coupled equations, the Raman polarization is then given by  $ P_{\rm R}= \left[\alpha_{11}\rho_{11}+\alpha_{22}\rho_{22}+ \alpha_{12}  \left(\rho_{12}+\rho_{12}^{*} \right)\right]  N_{0}E$. For weak Raman excitation, $ \rho_{11}\approx 1 $ and $ \rho_{22}\approx 0 $, i.e. the second term in $ P_{\rm R} $ can be neglected, while the first term increases the linear refractive index of the medium by a fixed amount.

Using Maxwell and Bloch equations and applying the SVEA, one can derive the following set of normalized coupled equations that govern pulse propagation in HC-PCFs filled by Raman-active gases,
\begin{equation}
\begin{array}{l}
\left[ i\partial_{\xi}+\displaystyle\sum_{m=2}\dfrac{i^{m}z_{0}}{t_{0}^{m}m!}\beta_{m}\partial_{t}^{m}+|\psi|^{2} +\dfrac{z_{0}}{z_{\mathrm{R}}}\,\mathrm{Re}\left(\rho_{12} \right)\right] \psi  =0,  \\
\partial_{\tau} w + \dfrac{\left( w+1\right)t_{0} }{T_{1}} =-4\, \mu\, w\, \mathrm{Im}\left(\rho_{12} \right)\, \left|\psi\right|^{2},\\ 
\left[ \partial_{\tau}  + \dfrac{t_{0}}{T_{2}}-i\delta\right]\rho_{12} =i\mu\, w\, \left|\psi\right|^{2},
\end{array}
\label{eq1} 
\end{equation}
where  weak Raman excitation is assumed, $ z_{\mathrm{R}}= c\,\epsilon_{0}/\left(\alpha_{12}\,N_{0}\,\omega_{0} \right)  $ is the nonlinear Raman length, $ \mu=P_{0}/P_{1} $, $P_{0}=A_{0}^{2}$, $ P_{1}=2\hbar\, c\,\epsilon_{0} A_{\mathrm{eff}} /\left(\alpha_{12}\,t_{0} \right) $,  $ \delta= \omega_{\rm R}t_{0}$, and  Re and Im represent the real and imaginary parts.

For femtosecond pulses, the relaxation times of the population inversion ($T_{1}$) and the coherence ($T_{2}$) can be safely neglected, since they are of the order of hundreds of picoseconds or more. For instance, $T_{1}\approx$ 20 ns and $T_{2}=$ 433 ps for excited rotational Raman in molecular hydrogen under gas pressure 7 bar at room temperature \cite{Belli15,Bischel86}. We have found also that the population inversion is almost unchanged from its initial value for pulses with energies in the order of few $ \mu $J, i.e. $ w\left(\tau \right)\approx w\left(-\infty \right)=-1 $. The set of the governing equations Eq. (\ref{eq1}) can be reduced to a single generalized nonlinear Schr\"{o}dinger equation,
\begin{equation}
i\partial_{\xi}\psi+i\frac{\beta_{1}z_{0}}{t_{0}}\partial_{\tau}\psi+\frac{1}{2}\partial_{\tau}^{2}\psi+|\psi|^{2}\psi   +R\left(\tau \right)\psi=0,
\label{eq2}
\end{equation}
where $ R\left(\tau \right)=\kappa \int_{-\infty}^{\tau}\sin\left[ \delta\left( \tau-\tau'\right) \right]  \left|\psi\left( \tau'\right) \right|^{2} d\tau'$ is the resulting Raman convolution, and $ \kappa=\mu z_{0}/z_{\mathrm{R}} $ is the ratio between the Raman and the Kerr nonlinearities. Pumping in the deep anomalous dispersion regime ($\beta_{2}<0$), it is assumed that higher-order dispersion coefficients $ \beta_{m>2} $ can be neglected. For ultrashort pulses with durations $t_{0}\ll1/\omega_{\rm R}$, $\sin\left[ \delta\left( \tau-\tau'\right) \right] $ can be expanded around the temporal location of the pulse peak by using the Taylor expansion. For instance, a fundamental soliton with amplitude $ N $ and centered at $ \tau=0 $ will induce a Raman contribution in the form of $ R\left(\tau \right)\approx \kappa N\sin\left( \delta\tau\right) \left [ 1+ \,\mathrm{tanh} \left(N\tau\right)  \right] $ at the zeroth-order Taylor approximation. This soliton will generate a retarded sinusoidal Raman polarization that can impact the dynamics of the other trailing probe pulse lagging behind it. On the other hand, for $t_{0}\gg1/\omega_{\rm R}$, $R\left(\tau \right)\approx \gamma_{R}\left|\psi\left( \tau\right) \right|^{2}$ with $ \gamma_{R}= \kappa/\delta$, i.e the Raman nonlinearity is considered to be instantaneous. So, the Raman contribution would induce an effective Kerr nonlinearity that is significant, and can compete directly with the intrinsic Kerr nonlinearity of the gas \cite{Belli15,Bartels03}.

\begin{figure}
\includegraphics[width=8.6cm]{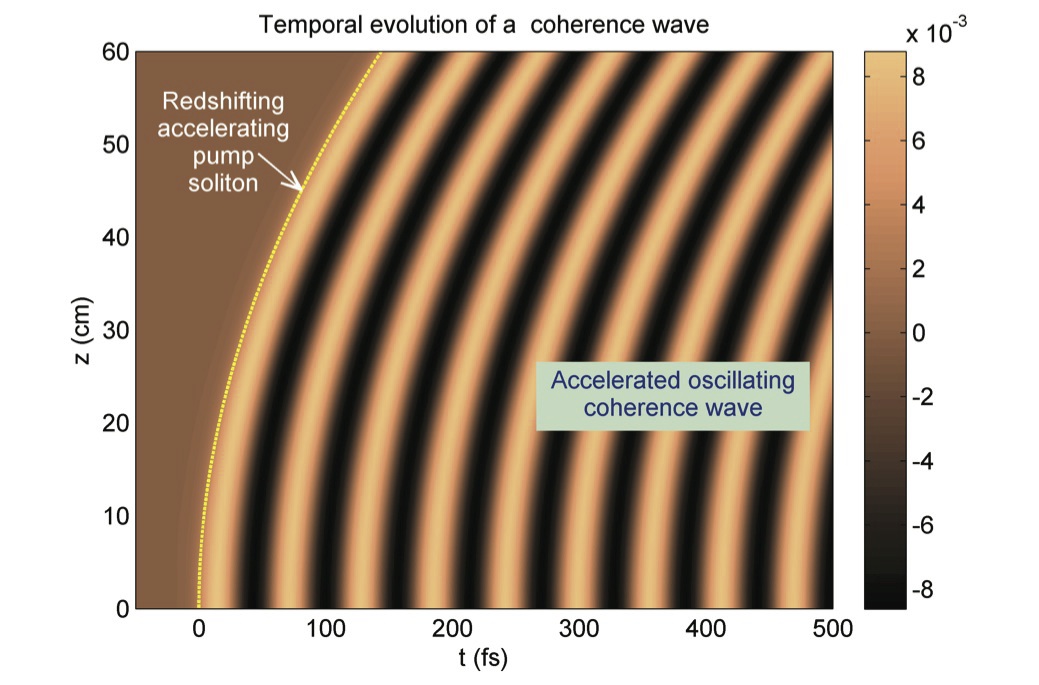}  
\caption{Temporal evolution of an accelerated oscillating Raman polarization with period $ \Lambda=56.7 $ fs induced by a propagating fundamental soliton with an amplitude $N_{1}=1.33$, a central wavelength $1064$ nm, and a FWHM $15$ fs in a H$ _{2} $-filled HC-PCF with a Kagome-lattice cross section, a flat-to-flat core diameter 18 $ \mu $m, a  zero dispersion wavelength 413 nm, a gas pressure $7$ bar, and a rotational Raman frequency $ \omega_{\rm R} = 17.6 $ THz. The dashed yellow line represents the temporal evolution of the soliton that excites the coherence wave. The simulation parameters are $ \gamma=7.07 \times 10 ^{-6} $ W$ ^{-1} $m$ ^{-1} $, $ \beta_{2}=-3425.5 $ fs$ ^{2} $/m, $ A_{\mathrm{eff}}=134\,\mu $m$ ^{2} $, $ \alpha_{12}=0.8 \times 10^{-41} $ C m$^{2}$/V \cite{Belli15,Weber94}, and $ t_{0}= $  11.34 fs. The parameter $ \gamma $ is calculated using the nonlinear susceptibility of H$ _{2} $ \cite{Mizrahi85}. For these parameters, we have found that higher-order dispersion and self steepening effects have a weak influence on the soliton dynamics.
\label{Fig3-1}}
\end{figure}

In the following, we will exploit the long Raman coherence induced by an ultrashort pulse in controlling pulse dynamics, see Fig. \ref{Fig3-1}. We will study the propagation of two successive pulses separated by a delay $\ll T_{1}, T_{2}  $ in the deep anomalous dispersion regime. The two pulses are assumed to have the same frequency, hence, they will propagate with the same group velocity, and experiencing the same dispersion. The leading pulse is an ultrashort strong `pump' pulse $ \psi_{1}$ with $t_{0}\ll1/\omega_{\rm R}$. In this case, Eq. (\ref{eq2}) can be used to determine the pump solution by replacing $ \psi $ by $ \psi_{1} $. For weak Raman nonlinearity, the solution of Eq. (\ref{eq2}) can be assumed to be a fundamental soliton that is perturbed by the Raman polarization, i.e. $ \psi_{1}\left( \xi,\tau\right)=N_{1}\,\mathrm{sech} \left[N_{1} \left(\tau-u_{1}\xi-\bar{\tau}_{1}\left(\xi \right) \right) \right] \exp\left[-i\Omega_{1}\left(\xi \right)\left(\tau-u_{1}\xi\right)\right] $, where $u_{1}=\beta_{11}z_{0}/t_{0}$,  $\beta_{11}$ is the first-order dispersion coefficient of the pump, $N_{1} $, $ \Omega_{1} $, and $\bar{\tau}_{1}  $  are the soliton amplitude, central frequency, and temporal location of the peak maximum, respectively. Assuming that we launch this soliton as a pump with $\bar{\tau}_{1}\left(0 \right)=0$, and using the variational perturbation method \cite{Agrawal07}, we have found that this soliton is linearly redshifting in the frequency domain with rate $ g_{1} =\frac{1}{2}\kappa\pi\delta^{2} \mathrm{csch}\left( \pi\delta/2N_{1}\right)$, and decelerating in the time domain, i.e. $ \Omega_{1} = - g_{1}\, \xi,$ and $ \bar{\tau}_{1} = g_{1}\, \xi^{2}/2 $. In the case $t_{0}<1/\omega_{\rm R}$, we have found that a factor of  $\approx\frac{1}{2}$ might be used to correct the overestimated value of $ g_{1} $, resulting from using the zeroth-order Taylor approximation. The treatments of the dynamics of the trailing pulse `probe' $ \psi_{2}$ are presented in Sec. \ref{Sec3-1} and \ref{Sec3-2}, when it is a weak long pulse and strong ultrashort soliton, respectively  \cite{Saleh15a,Saleh15b}. The study of the pump-probe dynamics will prove crucial in the future understanding of the essential building blocks of supercontinuum generation in Raman-active gases when excited by ultrashort pulses \cite{Belli15,Saleh15c}.

\subsection{Weak probe evolution}\label{Sec3-1}
When a second weak probe pulse is sent after the leading pump soliton, the probe evolution is ruled by the following equation:
\begin{equation}
 i\partial_{\xi}\psi_{2}+i u_{2}\partial_{\tau}\psi_{2}+\dfrac{1}{2m}\partial_{\tau}^{2}\psi_{2} +2\kappa N_{1}\sin\left( \delta\tilde{\tau}\right) \psi_{2}  = 0 ,     
\label{eq3}
\end{equation}
where $ u_{2}= \beta_{12} z_{0}/t_{0}  $, $ m=\left|\beta_{21}\right|/\left|\beta_{22}\right| $, $ \beta_{1j} $ and $ \beta_{2j} $ are the first and the second order dispersion coefficients of the $ j^{\mathrm{th}} $ pulse with $ j=1,2 $. Going to the reference frame of the leading decelerating soliton, $ \tilde{\tau}=\tau-u_{1}\xi-g_{1}\xi^{2}/2 $, and applying a generalized form of the Gagnon-B\'{e}langer phase transformation \cite{Gagnon90} $ \psi_{2}\left( \xi,\tilde{\tau}\right) =\phi\left(\xi,\tilde{\tau} \right) \exp\left[i\tilde{\tau}\left(g_{1}\xi +u_{1}-u_{2} \right) +i \left(g_{1}\xi+u_{1}-u_{2} \right)^{3} /6g_{1}\right]  $, Eq. (\ref{eq3}) becomes \cite{Saleh15a}
\begin{equation}
i\partial_{\xi}\phi=-\frac{1}{2m}\partial_{\tilde{\tau}}^{2}\phi  + \left[-2\kappa N_{1}\sin \left(\delta\tilde{\tau} \right)+g_{1}\tilde{\tau}\right] \phi.
\label{eq4}
\end{equation}
This equation is the {\em exact analogue} of the time-dependent Schr\"{o}dinger equation of an electron in a periodic crystal in the presence of an external electric field. In Eq. (\ref{eq4}) time and space are swapped with respect to the condensed matter physics system, as usual in optics, and we deal with a spatial-dependent Schr\"{o}dinger equation of a single particle `probe' with mass $ m $ in a \textit{temporal crystal} with a periodic potential $ U= -2\kappa N_{1}\sin \left(\delta\tilde{\tau} \right) $ in the presence of a constant force $ -g_{1} $ in the positive-delay direction.  The leading soliton excites a sinusoidal Raman oscillation that forms a periodic structure in the reference frame of the soliton, as shown in Fig. \ref{Fig3-1}. Due to soliton acceleration induced by the strong spectral redshift, a constant force is applied on this structure.  Substituting $ \phi\left(\xi,\tilde{\tau} \right) = f\left(\tilde{\tau}\right) \exp\left(  iq\xi\right)  $, Eq. (\ref{eq4}) becomes an eigenvalue problem with eigenfunctions $ f $, and eigenvalues $ -q $. The modes of this equation are the Wannier functions \cite{Wannier60} that can exhibit Bloch oscillations \cite{Bloch28}, intrawell oscillations \cite{Bouchard95}, and Zener tunneling \cite{Zener34} due to the applied force. 

\begin{figure}
\includegraphics[width=8.6cm]{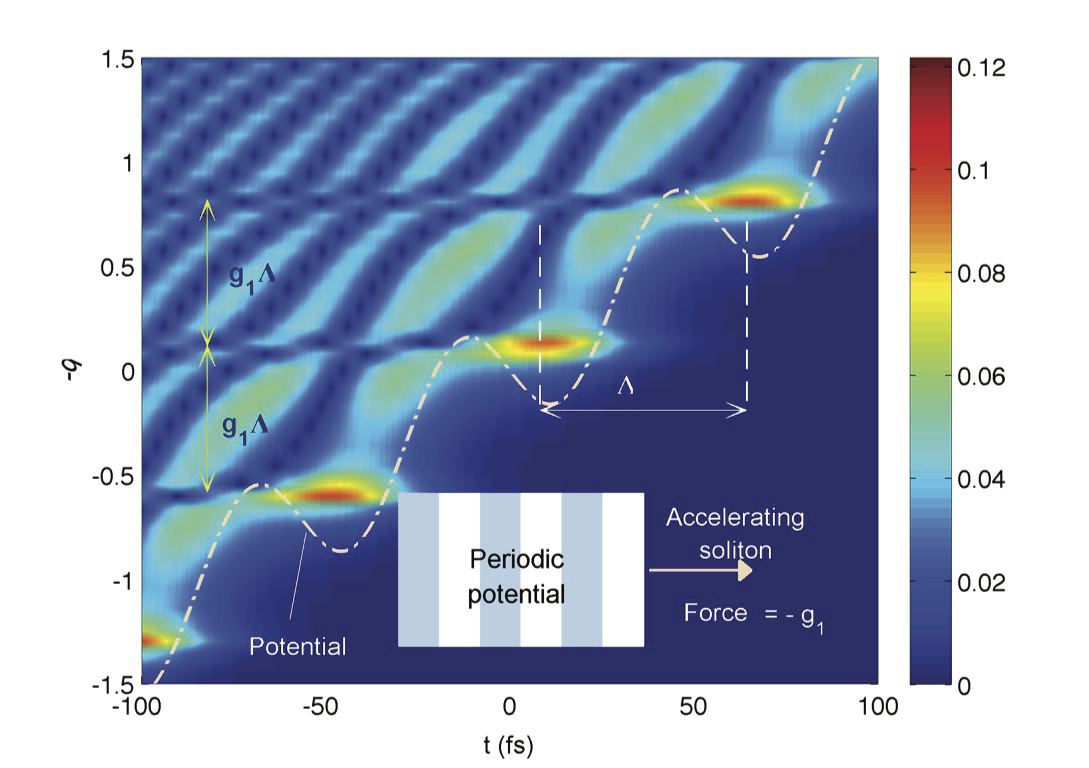} 
\caption{A portion of the absolute eigenstates of a Raman-induced temporal periodic crystals with a lattice constant $ \Lambda=56.7 $ fs in the presence of a force with magnitude $ g_{1}=0.1408 $ in the positive-delay direction. The vertical axis represents the corresponding eigenvalues $ -q $. The dotted-dashed line is the potential under the applied force. Other simulation parameters are similar to that used in Fig. \ref{Fig3-1}.
\label{Fig3-2}}
\end{figure}

\paragraph*{Wannier-Stark ladder---}
Consider the propagation of an ultrashort soliton with FWHM 15 fs in the deep anomalous dispersion regime of a H$ _{2} $-filled HC-PCF with a Kagome lattice.  Exciting the rotational Raman shift frequency in the fiber via this soliton will induce a long-lived trailing {\em temporal periodic crystal} with a lattice constant $ \Lambda= 56.7$ fs, see Fig. \ref{Fig3-1}, corresponding to the time required by the H$  _{2}$ molecule to complete one cycle of rotation. In the absence of the applied force, the solutions are the Bloch modes, while in the presence of the applied force, the periodic potential is tilted, and the eigenstates of the system are the Wannier functions portrayed as a  2D color plot in Fig. \ref{Fig3-2}, where the horizontal axis is the time and the vertical axis is the corresponding eigenvalue. These functions are modified Airy beams that have strong or weak oscillating decaying tails. After an eigenvalue step $ g_{1}\Lambda  $, the eigenstates are repeated, but shifted by $ \Lambda $, forming the Wannier-Stark ladder, well-known in condensed matter physics. As shown, each potential minimum can allow a single localized state with very weak tails. A large number of delocalized modes with long and strong tails exist between the localized states.  

\begin{figure}
\includegraphics[width=8.6cm]{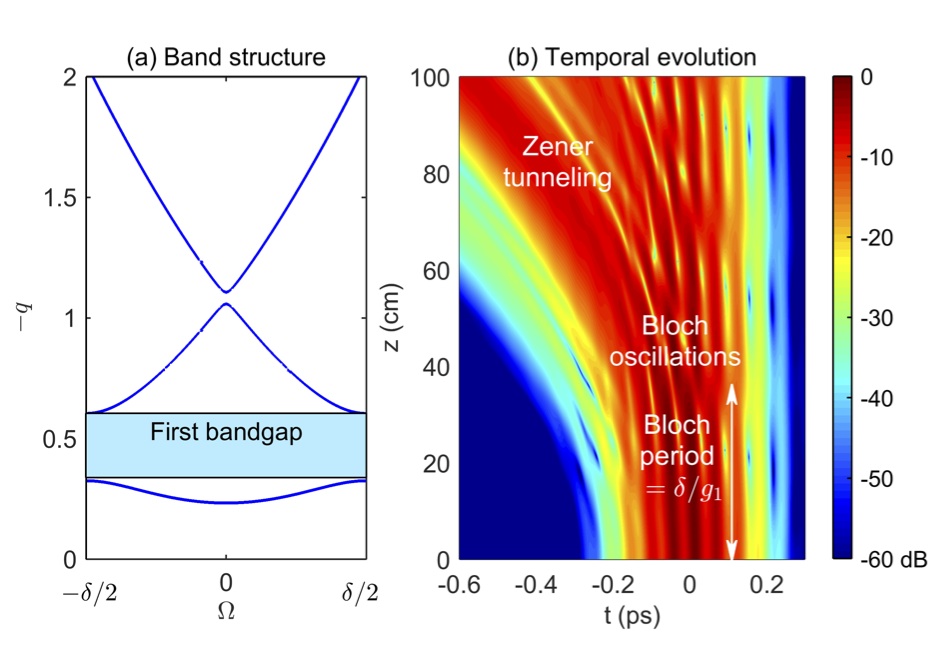} 
\caption{(a) Bandstructure of the temporal crystal induced by the leading ultrashort soliton propagating in the H$ _{2} $-filled Kagome-lattice HC-PCF with $ m=1 $ in the absence of the applied force. (b) Temporal evolution of a weak probe in the accelerated periodic temporal crystal. The probe initial temporal profile is a Gaussian pulse with FWHM 133.6 fs superimposed on the first Bloch mode of the periodic crystal in the absence of the applied force. The contour plot is given in a logarithmic scale and truncated at -60 dB. Other simulation parameters are similar to that used in Fig. \ref{Fig3-1}.
\label{Fig3-3}}
\end{figure}

\paragraph*{Bloch oscillations and Zener tunneling---}
An arbitrary weak probe following the soliton will be decomposed into the Wannier modes of the periodic temporal crystal. Due to beating between similar eigenstates in different potential wells, Bloch oscillations arise with a period $ \delta/g_{1} $, while beating between different eigenstates in the same potential minimum can result in intrawell oscillations.  In our case we did not observe in the simulations the latter kind of beating, since only a single eigenstate is allowed within each well. Interference between modes lying between different wells are responsible for Zener tunneling that allows transitions between different wells (or bands). In the absence of the applied force ($g_1=0$), the band structure of the periodic medium can be constructed by plotting the propagation constants of the Bloch modes over the first Brillouin zone $ \left[-\delta/2,\delta/2\right]  $, as shown in Fig. \ref{Fig3-3}(a). Zener tunneling occurs when a particle transits from the lowest band to the next-higher band. The evolution of a delayed probe in the form of the first Bloch mode inside a H$ _{2} $-filled HC-PCF under the influence of the pump-induced temporal periodic crystal, is depicted in Fig. \ref{Fig3-3}(b). Portions of the probe are localized in different temporal wells. Bloch oscillations are also shown with period 34.7 cm, which correspond to the beating between localized modes in adjacent wells. After each half of this period, an accelerated radiation to the left due to Zener tunneling is also emitted. Zener tunneling is dominant, and Bloch oscillations are weak, because the potential wells are relatively far from each other. The overlapping between the localized modes are small, consistent with the shallowness of the first band in the periodic limit (absence of the applied force). 

\subsection{Strong probe evolution}\label{Sec3-2} 
We now study the case when the delayed pulse is another strong fundamental soliton, rather than a weak probe pulse as in \cite{Saleh15a}. The governing equation for this strong `probe' is given by \cite{Saleh15b}
\begin{equation}
i\partial_{\xi}\psi_{2}+\frac{1}{2}\partial_{\tau}^{2}\psi_{2}+|\psi_{2}|^{2}\psi_{2}+R_{2}\left(\tau \right)\psi_{2}=0.
\end{equation}
For weak Raman nonlinearities, the solution of this equation is another perturbed fundamental soliton,  $ \psi_{2}\left( \xi,\tau\right)=N_{2}\,\mathrm{sech} \left[N_{2} \left(\tau-\bar{\tau}_{2}\left(\xi \right) \right) \right] \exp\left[-i\Omega_{2}\left(\xi \right)\tau\right] $ where  $N_{2} $, $ \Omega_{2}$, and $\bar{\tau}_{2} $  are the second soliton amplitude, central frequency, and temporal location of the peak maximum, respectively.  When the soliton duration $ \ll 1/\delta $, its Raman response function can be approximated by using a Taylor expansion as \cite{Saleh15a},
\begin{equation}
R_{2}\left(\tau \right)\approx \kappa\sum_{l=1,2 } N_{l} \sin\left[ \delta\left(\tau-\bar{\tau}_{l} \right)  \right] \left\lbrace  1+ \mathrm{tanh} \left[  N_{l} \left(\tau-\bar{\tau}_{l} \right) \right]  \right\rbrace.
\end{equation}
The superposition of the induced-Raman effects by the two solitons will affect the trailing soliton dynamics.

Adopting the variational perturbation method to understand how Raman nonlinearities can affect the pulse dynamics \cite{Agrawal07}, we have derived a set of coupled governing equations that determine the evolution of each soliton parameters \cite{Saleh15b}, the solutions of which are
\begin{equation}
\begin{array}{ll}
\Omega_{1} &= - g_{1}\, \xi,\\ 
\bar{\tau}_{1} &= g_{1}\, \xi^{2}/2,\\ 
\Omega_{2} &= - g_{2}\,\xi-g_{2}\dfrac{2N_{1}}{N_{2}}\displaystyle\int_{0}^{\xi} \cos\left[ \delta\left( \bar{\tau}_{2}-\bar{\tau}_{1}\right) \right]\,d\xi,\\
\bar{\tau}_{2}&=-\displaystyle\int_{0}^{\xi}\Omega_{2}\,d\xi,
\end{array} 
\label{eq5}
\end{equation}
where $ g_{j}=\frac{1}{2}\kappa\pi\delta^{2} \mathrm{csch}\left( \pi\delta/2N_{j}\right)  $. The first (leading) soliton will always linearly redshift in the frequency domain with rate $ g_{1} $, and decelerate in the time domain. Whereas for the second (trailing) soliton, its dynamics depends on two different components: (i) its own (self) component that will lead to a linear redshift similar to the leading soliton, with rate $ g_{2} $; (ii) a cross component representing the effect of the first soliton on the second soliton. The latter component is proportional to the ratio between their amplitudes and the cosine of the time difference between them. Since the cosine term varies between positive and negative values, the dynamics of the second soliton can switch back and forth between redshift and blueshift in the frequency domain or deceleration and acceleration in the time domain. This analytical model shows a very good agreement with the numerical model provided that the assumption of the soliton durations $ \ll 1/\delta $ is satisfied. This method can be extended easily to the case of more than two solitons.

\begin{figure}
\centerline{\includegraphics[width=8.6cm]{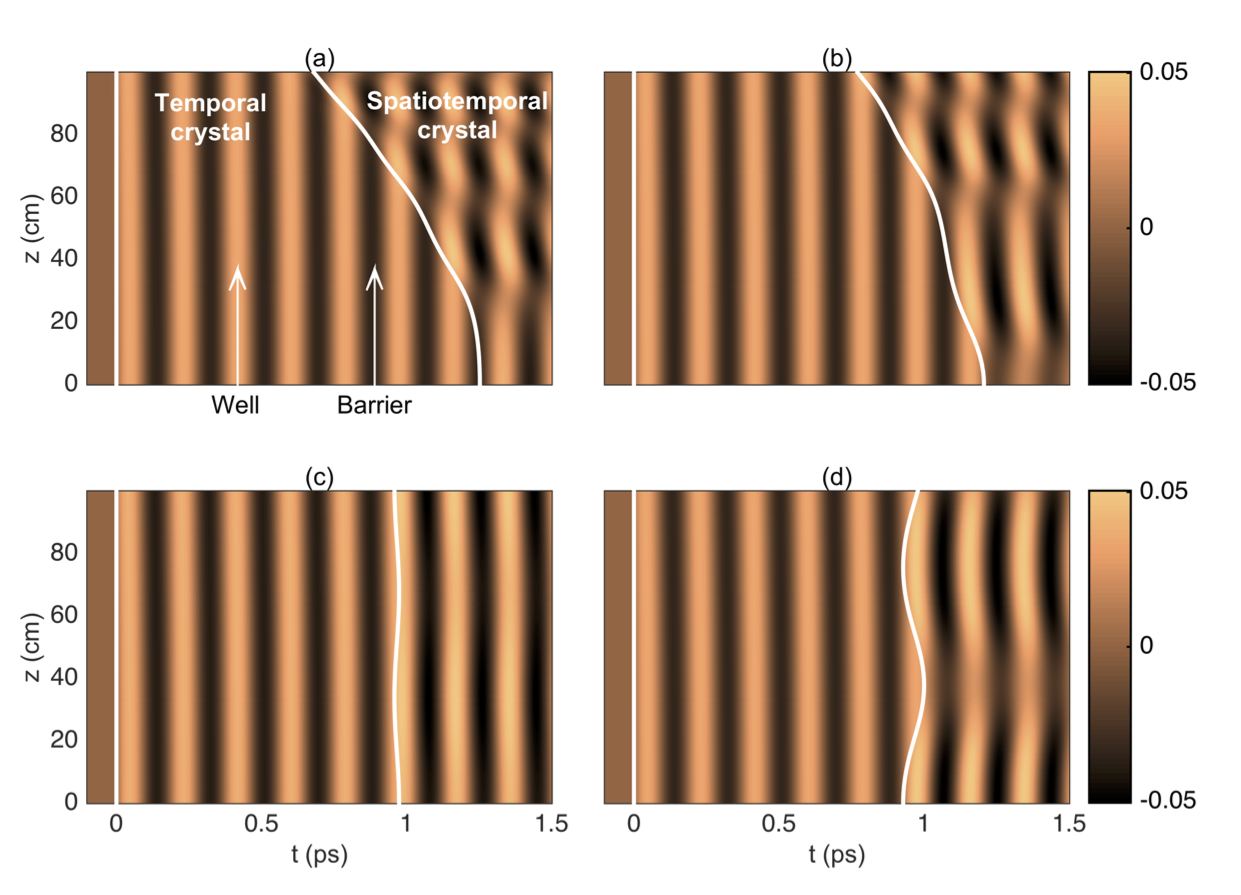}}
\caption{(Color online). Raman polarization induced by two fundamental solitons propagating in a gas-filled HC-PCF in a reference frame moving with the leading soliton. White solid lines represent the temporal evolution of the two solitons with normalized amplitudes $ N_{1} = 2.5 $, $ N_{2}=  1.25$ ( corresponding to full width half maximum (FWHM) 8 fs, 16 fs, respectively), using the analytical prediction Eq. (\ref{eq5}). The second soliton is launched at (a) $ \bar{\tau}_{2}\left(0 \right) =6.75\bar{\Lambda}$, (b) $ \bar{\tau}_{2}\left(0 \right) =6.5\bar{\Lambda}$, (c) $ \bar{\tau}_{2}\left(0 \right) =5.25\bar{\Lambda}$, and (d) $ \bar{\tau}_{2}\left(0 \right) =5\bar{\Lambda}$, where $ \bar{\Lambda}=2\pi/\delta $,  $ \delta=0.5 $  (equivalent to  a Raman-mode oscillation with period 185 fs, such as in deuterium \cite{Burzo07}), and $ \kappa=0.16 $.
\label{Fig3-4}}
\end{figure}

Figure \ref{Fig3-4} shows four special cases of the temporal evolution of the second soliton superimposed on the total induced-Raman polarization in a reference frame moving with the leading soliton $ \tilde{\tau}=\tau-g_{1}\xi^{2}/2 $. $ N_{2}$ is chosen smaller than $ N_{1} $ so that the cross component is comparable to the self component in Eq. (\ref{eq5}). The trailing soliton can be treated as  a  particle in a moving periodic potential. As we are operating in the anomalous dispersion regime, the positive (negative) variation of the refractive index represents a potential well (barrier). Thus, the periodic modulation of the refractive index corresponds to a sequence of alternative wells and barriers. Based on the initial time delay between the two solitons $ \Delta\bar{\tau}_{i} $, the dynamics of the second soliton  behaves differently. Also, the uniformity of the temporal crystal will be modified along the direction of propagation, resulting in a spatiotemporal modulation of the refractive index,  i.e. a {\em spatiotemporal crystal}. Looking at Fig. \ref{Fig3-4}, launching the second soliton at (a) the top of a barrier or (b) at the right edge of a well, the second soliton will be able to overcome the barriers during propagation, and transported across the potential by the acting force to the left direction. The output spatiotemporal crystals are chirped along the direction of propagation in these cases. Interestingly, the second soliton in (b) experiences a net maximum self-frequency blueshift of 9.12 THz $ \equiv 51.4 $ nm after 10 cm of propagation, before it is redshifted. Launching the second soliton at (c) a potential minimum or (d) a left edge of a well, the second soliton will not be able to overcome the barriers in these cases, so it is trapped inside the well and will oscillate indefinitely. The amplitude of oscillation in (c) is very small, since the initial velocity of the soliton in this potential is zero. The soliton will oscillate in an asymmetric manner  as in (d) due to the modified potential beyond the second soliton peak as well as the acting force that is opposite to the initial velocity. The resulting spatiotemporal crystals have uniform periods along the direction of propagation in these cases.

\begin{figure}
\centerline{\includegraphics[width=8.6cm]{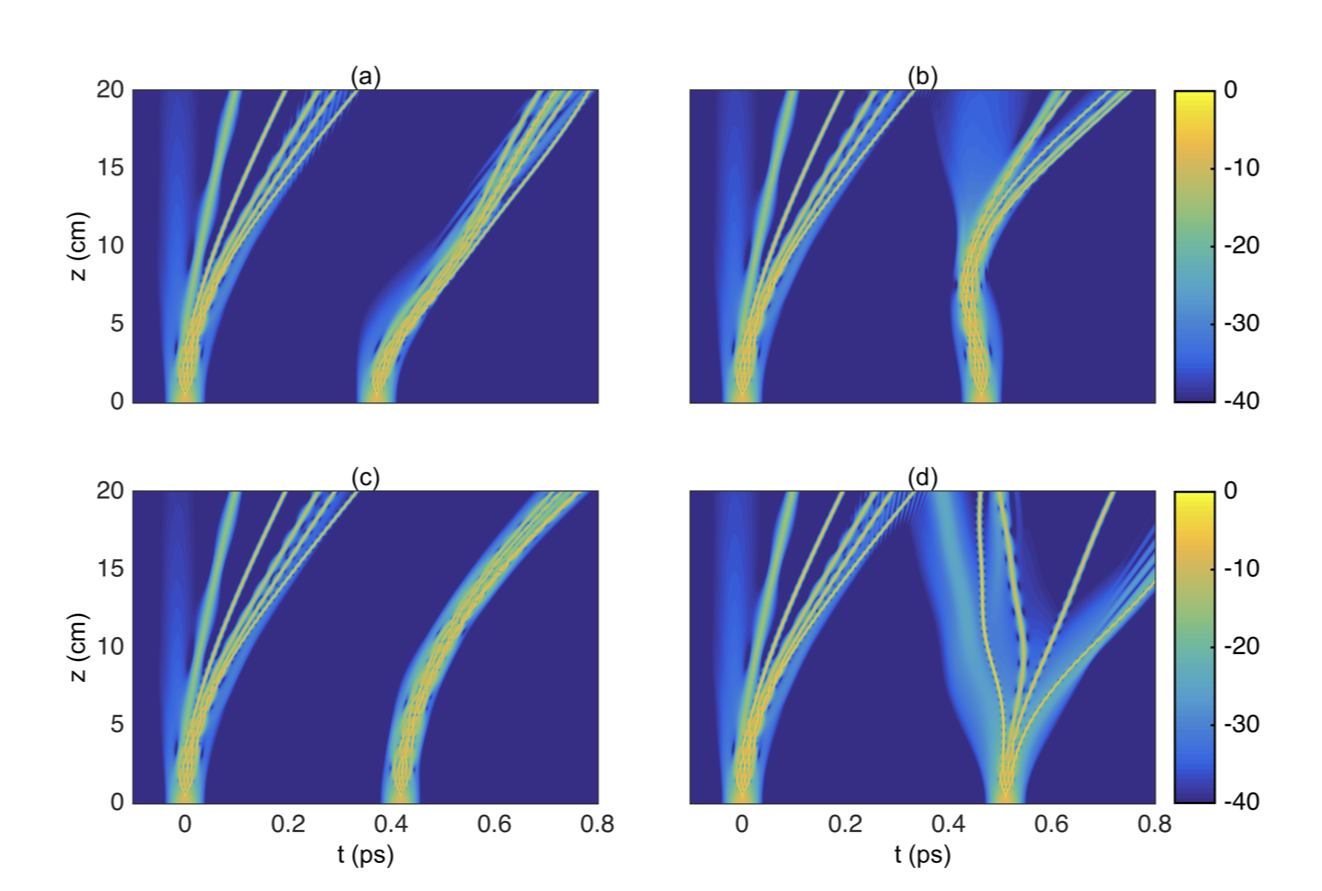}}
\caption{(Color online). Temporal evolution of two successive identical Gaussian pulses with profile $ N\exp\left(-\tau^{2}/2 \right)$ in the gas-filled HC-PCF described in Fig. \ref{Fig3-4}, with $ N=7 $, and FWHM = 25 fs. The first pulse is launched at $ \bar{\tau}_{1}\left(0 \right) =0$. The second pulse is launched at: (a) $ \bar{\tau}_{2}\left(0 \right) =2\bar{\Lambda}$, (b) $ \bar{\tau}_{2}\left(0 \right) =2.5\bar{\Lambda} $, (c) $\bar{\tau}_{2}\left(0 \right) =2.25\bar{\Lambda}$, and (d) $\bar{\tau}_{2}\left(0 \right) =2.75\bar{\Lambda} $.
\label{Fig3-5}}
\end{figure}

The temporal evolution of two successive ultrashort Gaussian pulses rather than fundamental solitons are depicted in Fig. \ref{Fig3-5} for different time delays. The two pulses have the same central frequencies and amplitudes, and the delay is again within the relaxation coherence time of the Raman-active gas. The two pulses will experience pulse compression and soliton fission processes. The `first' leading pulse excites the Raman polarization `potential' that will affect the `second' trailing pulse. The dynamics of the leading pulse is certainly independent of the time delay and will encounter a self-induced Raman redshift (deceleration). The trailing pulse dynamics is influenced by its self Raman-induced effect as well as the cross-accelerated Raman polarization effect of the leading pulse. In Fig. \ref{Fig3-5}, panel (a) shows the case when the self and cross components are working together, resulting in a strong delay in comparison to the first pulse. The situation when the self and cross components acts against each other is figured in panel (b), where initially the second pulse is nearly halted since the cross and self components cancel each other. When the second pulse is launched at one minima of the potential induced by the first pulse, the pulse is well-confined during propagation. Even after the pulse fission, the generated solitons are still traveling together, see panel (c). Launching at one potential-maxima, the dynamics of a tree-like behavior is obtained as shown in panel (d), where each soliton propagates in a different direction. In all these cases, the dynamics of each soliton depends on where exactly this soliton is born inside the total accelerated periodic potentials induced by other preceding solitons.

The above results will be of fundamental importance in the understanding of the building blocks of supercontinuum generation: the multitude of solitons propagating in the fiber influence each other in a well-defined way, defined by their intensities and their temporal separations. The condensed matter physics analogue effects will take place and they will determine the dynamics of each individual solitons in the supercontinuum process \cite{Belli15,Saleh15c}. This theory allows new nonlinear phenomena that are impossible to achieve in conventional solid-core optical fibers, and opens up new exciting venues for future discoveries.

\section{Ionization effect in gas-filled HC-PCFs}
\paragraph*{Ionization Models---} Photoionization is the physical process in which an electron is released and an ion is formed due to the interaction of a photon with an atom or a molecule.  The free electron density $ n_{e} $ is governed by the rate equation \cite{Sprangle02}
\begin{equation}
 \dfrac{\partial n_{e}}{\partial t}=\mathcal{W}\left( t\right)\left(n-n_{e} \right)  -\eta\, n_{e}- \beta_{r}n_{e}^{2}, \label{eqNe}
 \end{equation} 
where $\mathcal{W}  $ is the ionization rate, $ n $ is the total density of the atoms, and $ \eta $ and $ \beta_{r} $ are the electron attachments, and recombination rates, respectively. For pulses with pico-second duration or less, both $ \eta $ and $ \beta_{r} $ are negligible. Based on the so-called Keldysh parameter  $p_{\rm K}$ \cite{Keldysh64}, photoionization can take place by multiphoton absorption  $p_{\rm K}\gg 1$,  tunneling ionization  $p_{\rm K}\ll 1$ or both when $p_{\rm K}\sim 1$. Several models have been developed to determine the dependence of the  ionization rate on the electric field of the optical pulse, for instance, multi-photon ionization-based Keldysh-Faisal-Reiss model  \cite{Reiss90}, tunneling based Ammosov- Delone-Krainov method \cite{Ammosov86}, Perelomov, Popov, and Terent’ev (PPT) hybrid technique \cite{Perelomov66}, and Yudin-Ivanov model \cite{Yudin01},  which is a modification of the PPT technique. It has been shown experimentally that tunneling ionization is dominant over the multiphoton ionization in noble gases for pulses with intensities $ \sim 10^{14} $ W/m$ ^{2} $ \cite{Gibson90,Augst91}, which is our case of study. A review of these processes and models is found in Ref. \cite{Popov04}. In the tunneling regime, the time-averaged ionization rate is given by  \cite{Sprangle02}
\begin{equation}
\mathcal{W}(I)=d\,(I_{H}/I) ^{1/4}\, \exp[ -b\,(I_{H}/I)^{1/2}], \label{eqW}
\end{equation}
where $ d= 4\,\delta_{0}\,[3/\pi]^{1/2}\,[U_{I}/U_{H}]^{7/4} $, $b= 2/3\, [U_{I}/U_{H}]^{3/2} $, $ \delta_{0}=4.1\times 10^{16}$ Hz is the characteristic atomic frequency, $ U_{I} $ is the ionization energy of the gas, $ U_{H}\approx 13.6 $ eV is the ionization energy of atomic hydrogen, $ I_{H} =3.6 \times 10^{16} $ W/cm$^{2}$ and $I=\left| \psi\right|^{2}$ is the laser pulse intensity. This model is adequate for the analysis that is based on the evolution of the complex envelope of the pulse. Unfortunately, all these models are not straightforwardly amenable to analytical manipulation, because of the complex dependence on the pulse intensity. Equation (\ref{eqW}) predicts an ionization rate that is exponential-like for pulse intensities above a threshold value $ I_{\mathrm{th}} $ \cite{Saleh11a,Saleh11b}. Any pulse with an intensity $ I\gg I_{\mathrm{th}} $ will suffer  a strong ionization loss due to the absorption of photons in the plasma generation process. This limits the operating regime to near the ionization threshold  $ I_{\mathrm{th}} $ where the ionization loss is drastically reduced. A model of the ionization rate with linear dependence on the pulse intensity can be developed using the first-order Taylor series of Eq. (\ref{eqW}) \cite{Saleh11a}
\begin{equation}
\mathcal{W}\approx \tilde{\sigma}\left( I -I_{\mathrm{th}}\right)  \Theta \left( I -I_{\mathrm{th}}\right),
\end{equation}
where $ \tilde{\sigma} $ is a constant that is chosen to reproduce the physically observed value of the ionization threshold, and $ \Theta $ is a Heaviside function, introduced to cut the ionization rate to zero below the value of $ I_{\mathrm{th}} $.

Photoionization results in decreasing the refractive index of the medium by a factor proportional to the square of the plasma frequency $ \omega_{p} $, also it attenuates the pulse amplitude due to photon absorption. To include these effects Eq. (\ref{eqNLSE}) can be modified as \cite{Saleh11a}
\begin{equation}
i\partial_{z}A+\sum_{m=2}\dfrac{i^{m}}{m!}\beta_{m}\partial_{t}^{m}A+\gamma|A|^{2}A-\dfrac{\omega_{p}^{2}A}{2\omega_{0}c} +i\dfrac{A_{\mathrm{eff}}U_{I}}{2\left|A\right|^{2}} \partial_{t} n_{e}=0, \label{GNLSE2}
\end{equation}
where $ c $ is the speed of light, and $ A_{\mathrm{eff}} $ is the effective area. Using the split-step Fourier method, the pulse amplitude can be determined at each propagation step after computing the free electron density $ n_{e} $. After neglecting the electron attachments and recombination effects, Eq. (\ref{eqNe}) can be directly integrated analytically, and can then be substituted in Eq. (\ref{GNLSE2}) to have a single generalized NLSE to describe pulse propagation in an ionizing medium. Introducing the following rescalings and redefinitions:  $ \omega_{\mathrm{T}}^{2}=e^{2} n/\left[ \epsilon_{0}m_{\mathrm{e} }\right] $ is the maximum plasma frequency, $\phi= \frac{1}{2}k_{0}z_{0}\,[\omega_{\mathrm{p}}/\omega_{0}]^{2}$, $\phi_{\mathrm{T}}=\frac{1}{2}k_{0}z_{0}\,[\omega_{\mathrm{T}}/\omega_{0}]^{2}$, $\sigma=\tilde{\sigma}\,t_{0}/[ A_{\mathrm{eff}}\gamma_{\mathrm{K}}\,z_{0}]$, and $ \kappa= U_{I}\, \tilde{\sigma}\,\epsilon_{0}\,m_{\mathrm{e}}\, \omega_{0}^{2}/[k_{0}\,e^{2}]$.
In this case, Eqs. (\ref{eqW}-\ref{GNLSE2}) become \cite{Saleh11b}
\begin{equation}
i\partial_{\xi}\psi+\sum_{m=2}\frac{i^{m}z_{0}}{t_{0}^{m}m!}\beta_{m}\partial_{\tau}^{m} \psi+|\psi|^{2}\psi-\phi\psi+i\alpha\psi=0,\label{nGNLSE2}
\end{equation}
where $ \phi=\phi_{\mathrm{T}}\left[1-\mathrm{exp} \left( -\sigma\int_{-\infty}^{\tau}\Delta|\psi|^{2}\,\Theta(\Delta|\psi|^{2})\,d\tau'\right) \right]  $, $ \alpha=\kappa\,\left(\phi_{\mathrm{T}}-\phi \right)\,\Delta|\psi|^{2} \Theta \Delta|\psi|^{2}   $, and  $\Delta|\psi|^{2}=|\psi|^{2}-|\psi|^{2}_{\rm th}$,  $ |\psi|_{\mathrm{th}}^{2} =I_{\mathrm{th}}A_{\mathrm{eff}}$.

\subsection{Short pulse evolution} 
In order to extract useful analytical information from Eq. (\ref{nGNLSE2}), further simplifications are necessary. First, we assume operating in the deep anomalous regime. For pulses with maximum intensities just above the ionization threshold, also called {\em floating} pulses \cite{Saleh11a}, the ionization loss is not large and can be neglected as a first approximation. For such pulses, only a small portion of energy above the threshold intensity contributes to the creation of free electrons. Furthermore, the effect of the $\Theta$-function can be approximately determined via multiplying the cross-section parameter $\sigma$ by a factor $ \varepsilon $ that represents the ratio between the pulse energy contributing to plasma formation (the portion above the ionization threshold) and the total energy of the pulse. In this case, ionization can be treated as a perturbation of the solution of the NLSE, which is the fundamental soliton. The solution of Eq. (\ref{nGNLSE2}) can be written as $\psi(\xi,\tau)=N(\xi)\,{\mathrm{sech}}\left[N(\xi)(\tau-\bar{\tau}(\xi))\right]e^{-i\delta(\xi)\tau}$, where $\bar{\tau}$ is the temporal location of the soliton peak, and $\delta$ is the  pulse central-frequency shift. Applying the perturbation theory \cite{Agrawal07}, $\delta(\xi)=-g\,\xi  $, and $\bar{\tau}(\xi)=g\,\xi^{2}/2$, where  $g=-(2/3)\varepsilon\sigma\phi_{\mathrm{T}} N^{2}$ \cite{Saleh11a}. This shows that photoionization should lead to an absolutely remarkable {\em soliton self-frequency blueshift}. This blueshift is accompanied by a constant acceleration of the pulse in the time domain -- opposite to the Raman effect, which produces pulse deceleration \cite{Mitschke86}. 

\begin{figure}
\includegraphics[width=8.6cm]{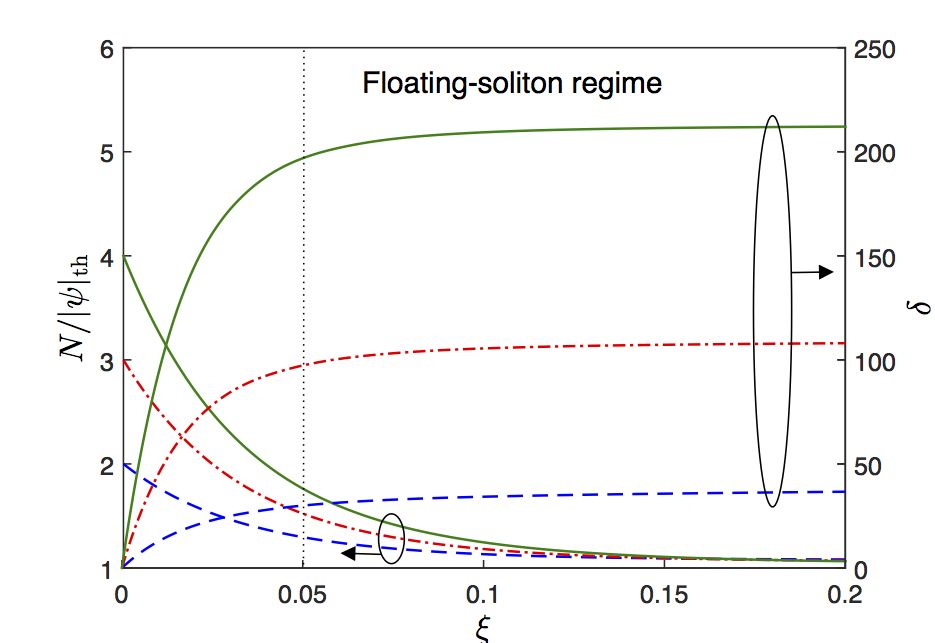}
\caption{(Color online). The spatial dependence of the soliton amplitude and frequency shift during a photoionization process for different initial pulse amplitudes.}
\label{Fig4-1}
\end{figure}

Including the effects of both the photoionization loss and the Heaviside function, the perturbation theory results in a set of two coupled differential equation that governs the spatial evolution of the soliton amplitude and central frequency,
\begin{equation}
\begin{array}{ll}
\dfrac{\partial N}{\partial\xi}= &-2\kappa\,\phi_{\mathrm{T}}\,\left(N\,\mathrm{tanh}\,\vartheta -|\psi|_{\mathrm{th}}^{2}\,T\right)\vspace{0.5mm}  \\
 \dfrac{\partial\delta}{\partial\xi}=& \sigma\phi_{\mathrm{T}} N^{2}\left[ \dfrac{2}{3}\,\mathrm{tanh}^{3}\vartheta + \dfrac{|\psi|_{\mathrm{th}}^{2}}{N^{2}}\left(\vartheta \mathrm{sech}^{2}\vartheta-\mathrm{tanh}\vartheta\right) \right],
\end{array}
\end{equation}
where $ \vartheta=NT $, and $ T\approx 1/N \mathrm{sech}^{-1} \left[|\psi|_{\mathrm{th}}/N\right] $ is the temporal position where the pulse amplitude exceeds the ionization threshold at $ \xi=0 $ \cite{Saleh11b}. Solving these equations numerically as shown in Fig. \ref{Fig4-1}, we found that pulses with initially large intensities $\left( N_{0}^{2} > |\psi|^{2}_{\mathrm{th}}\right)$ will experience a boosted self-frequency blueshift. However, the ionization loss suppresses the soliton intensity after a short propagation distance to the floating-soliton regime, where the soliton can propagate for a long propagation distance with a limited blueshift and negligible loss. The maximum frequency shift is achieved when the soliton intensity falls below the photoionization threshold.

\begin{figure}
\includegraphics[width=8.6cm]{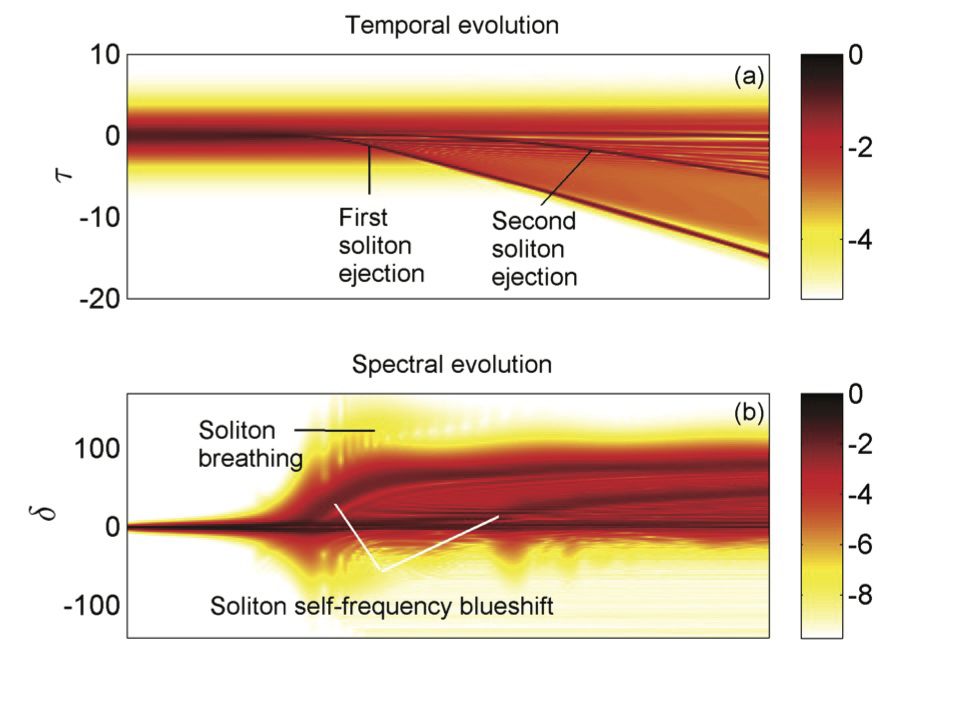}\vspace{-0.7cm}
\includegraphics[width=8.6cm]{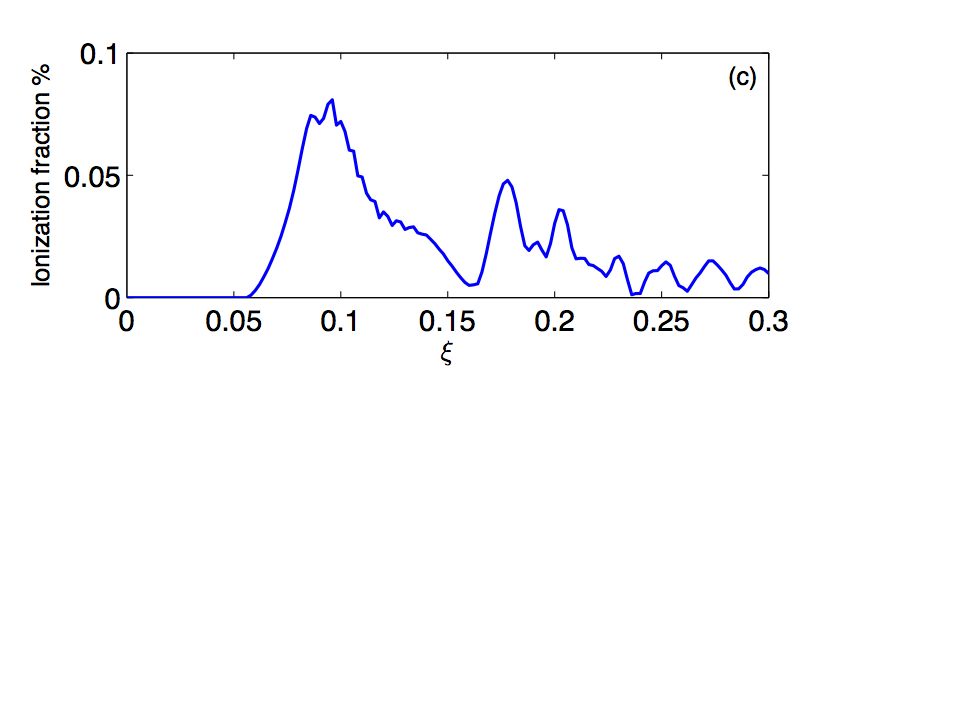}\vspace{-3 cm}
\caption{(Color online). Temporal (a) and spectral (b) evolution of an ultrashort pulse with in an Ar-filled HC-PCF. The temporal profile of the input pulse is $ N\,\mathrm{sech}\,\tau $, with $ N=8 $, $ t_{0}= 50 $ fs. The gas pressure is 5 bar.  Contour plots are given in a logarithmic scale. (c) Spatial dependence of the ionization fraction along the fiber.
\label{Fig4-2}}
\end{figure}

\begin{figure}
\includegraphics[width=8.6cm]{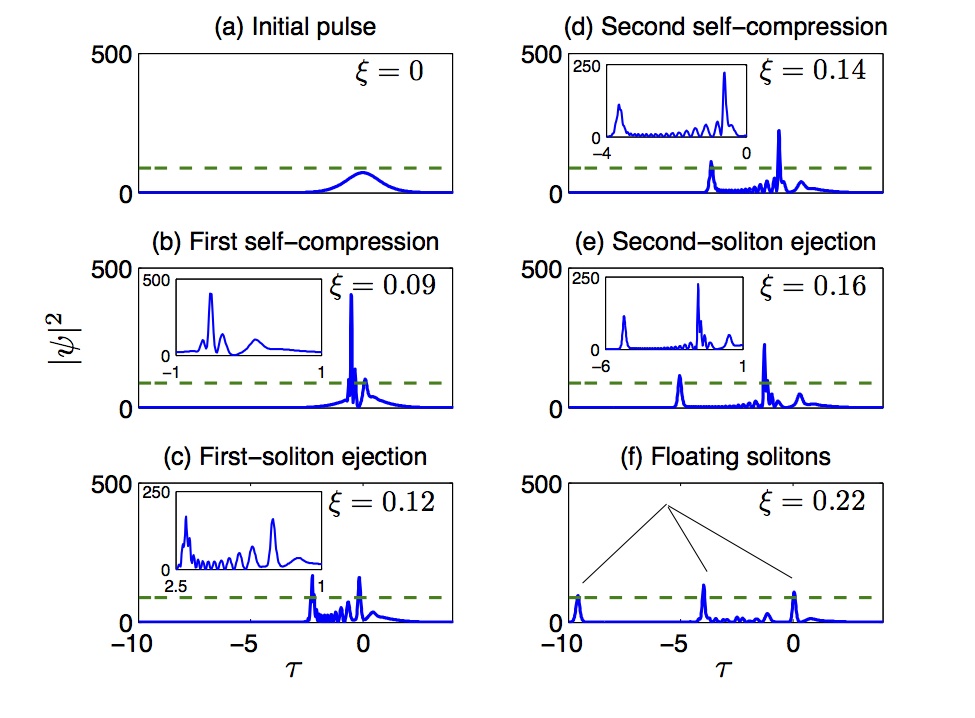}
\caption{(Color online). Intensity profile of an optical pulse  in the time domain at different positions, $ \xi $, inside an Ar-filled HC-PCF. The dashed red line represents the threshold intensity. The simulation parameters are similar to Fig. \ref{Fig4-2}. Each panel is titled by its main feature. Insets are enclosed in panels for better view and more details.
\label{Fig4-3}}
\end{figure}

To study the full dynamics of pulse propagation in fibers filled by an ionizing gas, Eq. (\ref{nGNLSE2}) should be solved numerically via the split-step Fourier method. The temporal and spectral evolution of a pulse, with an initial temporal profile $ N \mathrm{sech}\left(\tau \right)  $ and intensity less than the ionization threshold, are depicted in the panels (a,b) of Fig. \ref{Fig4-2}, respectively. Panel (c) shows the variation of the ionization fraction along the fiber. The pulse is pumped in the deep anomalous-dispersion regime of the fiber, and it undergoes self-compression. When the pulse intensity exceeds the threshold value, a certain amount of plasma is generated due to gas ionization, and a fundamental soliton is ejected from the input pulse. The soliton central frequency continues to shift towards the blue-side due to the energy received from the generated plasma. However, because of the concurrent ionization loss, the soliton intensity gradually attenuated to the regime where $ |\psi|^{2} \gtrapprox |\psi|^{2}_{\mathrm{th}} $.  Such pulses, the floating solitons, can propagate for considerably long distances with minimal attenuation and limited blueshift. When the soliton intensity goes below the ionization threshold, the blueshift process is ceased. A second ionization event accompanied by a second-soliton emission can take place by further self-compression of the input pulse based on its initial intensity. At the end, a train of floating solitons are generated. A clear representation for the pulse dynamics in the presence of plasma is shown in Fig. \ref{Fig4-3}, where the temporal profile of the pulse intensity $ |\psi|^{2} $ is plotted at selected positions inside the fiber. 

\begin{figure}
\includegraphics[width=8.6cm, height=7cm]{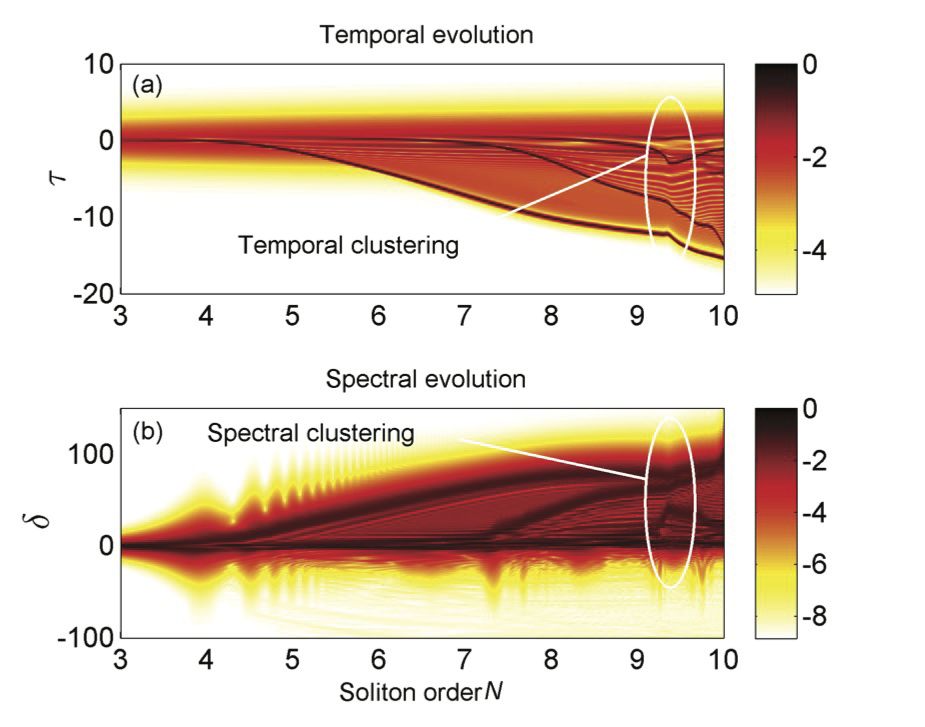}
\caption{(Color online). Temporal (a) and spectral (b) outputs of a pulse with temporal profile $ N\,\mathrm{sech}\,\tau $ after propagating inside an Ar-filled HC-PCF with length $\xi= 1/4 $ versus the soliton parameter $ N $. 
\label{Fig4-4}}
\end{figure}

\begin{figure}
\includegraphics[width=8.6cm]{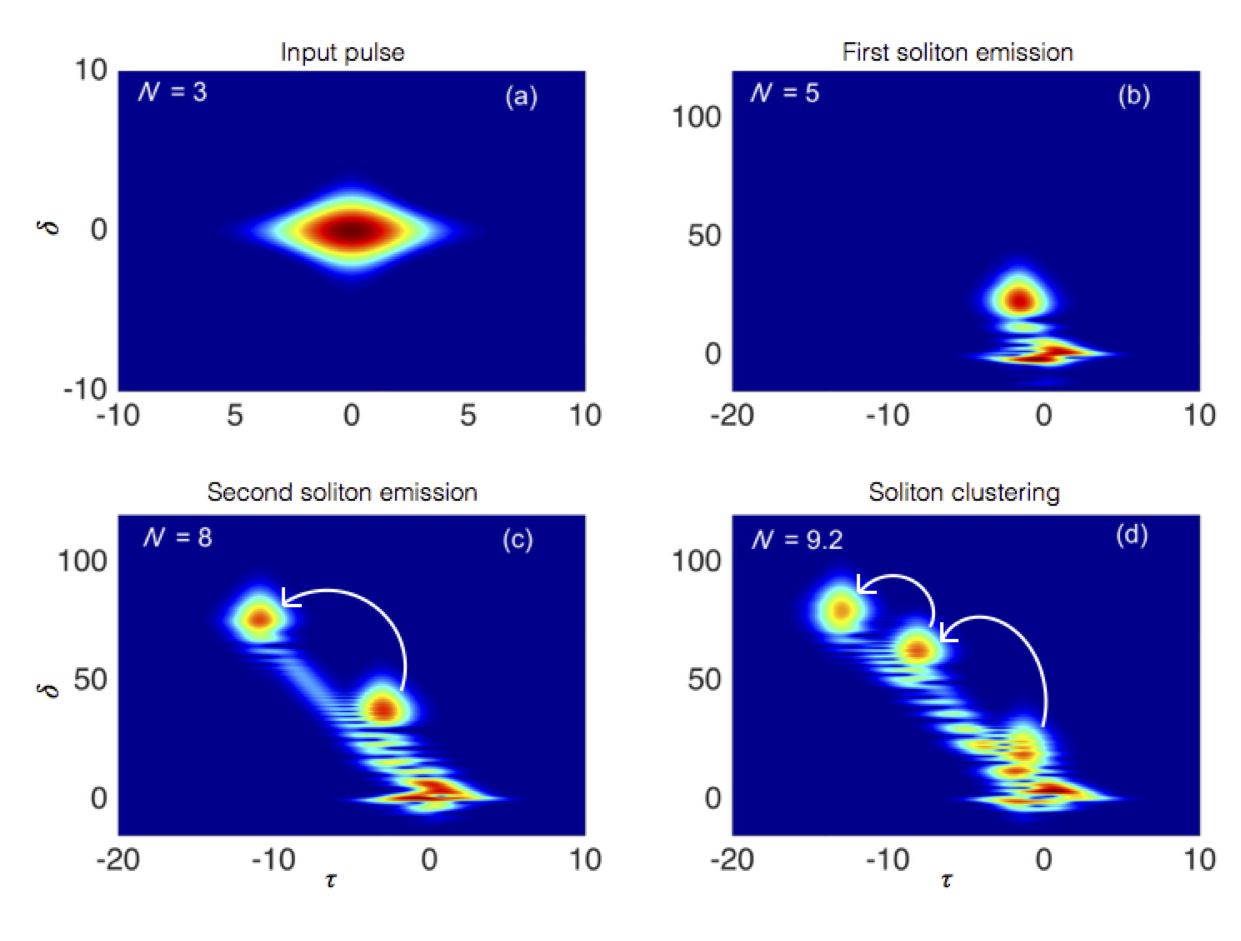}
\caption{(Color online). XFROG spectrograms for pulses with selected soliton parameter $ N $ in an increasing order. The simulation parameters are similar to Fig. \ref{Fig4-2}. (a) $ N=3,\, \xi=0 $. (b) $ N=5,\,  \xi= 1/4$. (c) $ N=8,\,  \xi= 1/4 $. (d) $ N=9.2,\,  \xi= 1/4 $. Each panel is titled by its main feature. White arrows show the movement of the solitons. \label{Fig4-5}}
\end{figure}

\paragraph*{Long-range non-local soliton forces and clustering ---} 
A non-local interaction between two successive solitons have been found when their temporal separation is shorter than the recombination time \cite{Saleh11a}, similar to the nonlinear interactions in Raman-active gases within the molecular coherence relaxation time presented in Sec. III. The leading soliton and its induced non-vanishing electron-density tail both co-propagate and accelerate towards the negative-delay direction. Within the recombination time, the trailing soliton will be affected by a force in the opposite direction due to the accelerated long electron-density tail. The acceleration of the trailing soliton acquires an exponentially decaying dependence on the amplitude of the leading soliton.  These  dynamics are featured in Fig. \ref{Fig4-4}, that shows the temporal and spectral dependence on the soliton parameter $ N $ assuming that the input pulse is $ N\,\mathrm{sech}\,\tau $. The scenario is as follows: As long as the intensity of the first-emitted soliton is above the threshold intensity, it prevents the ejection of a second soliton due to the presence of the opponent force of the non-local interaction. When the ionization loss ends the blueshifting process of the leading soliton, the trailing soliton can be ejected and recovers its expected acceleration and blueshift.  This allows the second soliton to catch up and cluster with the first soliton. In addition, the spectrum of the two solitons start to overlap and form spectral clustering. Similarly, when the first two solitons are very close to each other, their induced electron-density tail applies a combined force on the third soliton. The evolution of the cross-frequency-resolved optical gating (XFROG) spectrograms for pulses with different initial intensities are depicted in the panels of Fig. \ref{Fig4-5}, where (a) represents the input pulse; (b) and (c) shows the emission of the first and second solitons, respectively; and (d) depicts the temporal and spectral clustering of the first two solitons and the emission of a third soliton.

\subsection{Long-pulse evolution} 
It has been shown recently that when the gas is excited by relatively {\em long} pulses with ionizing intensities, new kinds of self-phase modulation (SPM) and modulational instability (MI) can emerge during propagation. Moreover, after the initial stage of instability is over, a `shower' of hundreds of solitons, each undergoing an ionization-induced self-frequency blueshift, pushes the supercontinuum spectrum towards shorter and shorter wavelengths. Such a {\em blueshifting} plasma-induced continuum has some similarities with the {\em redshifting} Raman-induced continuum driven by the Raman self-frequency shift in conventional solid-core fibers \cite{Islam89,GouveiaNeto89,Dianov89}, although the physical processes involved are dramatically different.

\begin{figure}
\includegraphics[width=8.6cm]{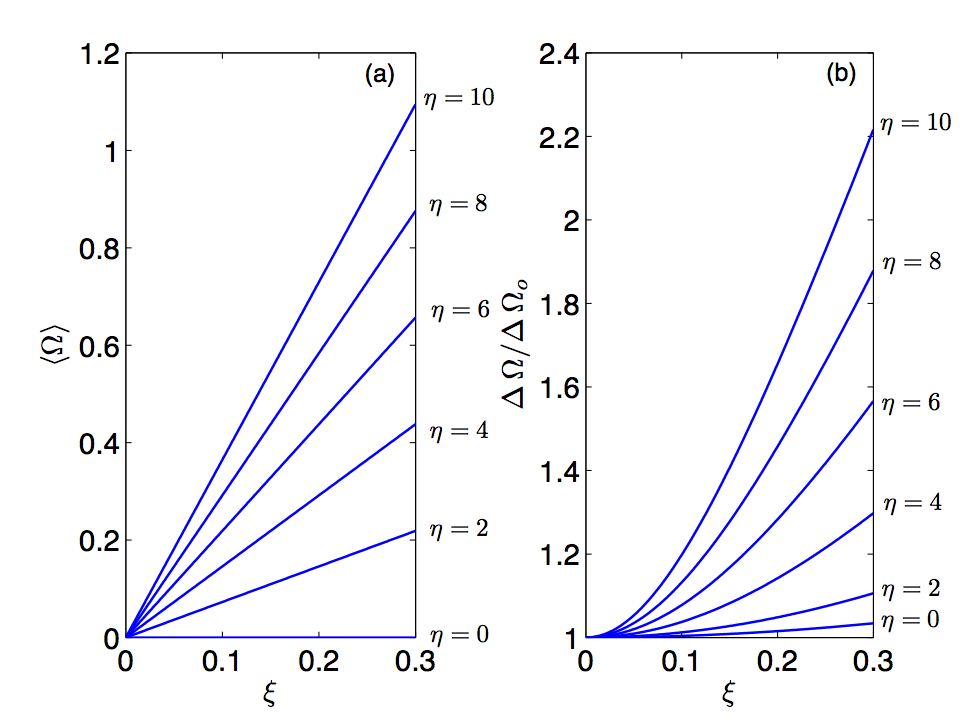}
\caption{(Color online). Spatial dependence of (a) the mean frequency $ \left\langle \Omega\right\rangle  $ and (b) the frequency standard deviation $ \Delta\Omega $ of a Gaussian pulse $  \exp \left(-\tau^{2}/2\,\tau_{0}^{2} \right) $ with $ \tau_{0}=2 $. The temporal position $ \mathrm{T}$ at which the pulse intensity can initiate photoionization is assumed to be equal to $\tau_{0} $. $ \Delta \Omega_{0} $ is the spectral width at $ \xi=0 $. \label{Fig4-6}}
\end{figure}

\paragraph*{Asymmetrical SPM---} Ionization-induced SPM can be studied analytically using Eq. (\ref{nGNLSE2}), in the  case of small dispersion and long input pulse durations the nonlinearity initially dominates over the group-velocity dispersion (GVD). Also by temporarily neglecting the losses (which do not change the qualitative picture, but only saturate the SPM spectrum after a certain distance), and assuming Gaussian pulse excitation, close forms of the spatial dependence of the mean frequency  $ \left\langle \Omega\right\rangle  $ and the standard deviation $   \left( \Delta\Omega \right) ^{2}$ of the output spectrum can be derived \cite{Saleh12}. For different values of $ \eta=\sigma\phi_{\mathrm{T}} $, which measures the ionization strength, panels (a,b) in Fig. \ref{Fig4-6} depict the spatial dependence of $ \left\langle \Omega\right\rangle  $ and $   \left( \Delta\Omega \right) ^{2}$ along the fiber. For $ \eta=0$, which corresponds to the absence of ionization, the mean frequency is always zero during propagation due to the well-known symmetric spectral broadening due to conventional SPM \cite{Agrawal07}.  As $ \eta $ increases, the plasma starts to build up inside the fiber. The mean frequency moves linearly towards the blue-side of the spectrum along the fiber due to the ionization-induced phase-modulation. This induces a strong and extremely asymmetric SPM, imbalanced towards the blue part of the spectrum. In a real case, the spectral broadening process is certainly limited by the unavoidable ionization and fiber losses.

\begin{figure}
\includegraphics[width=8.6cm]{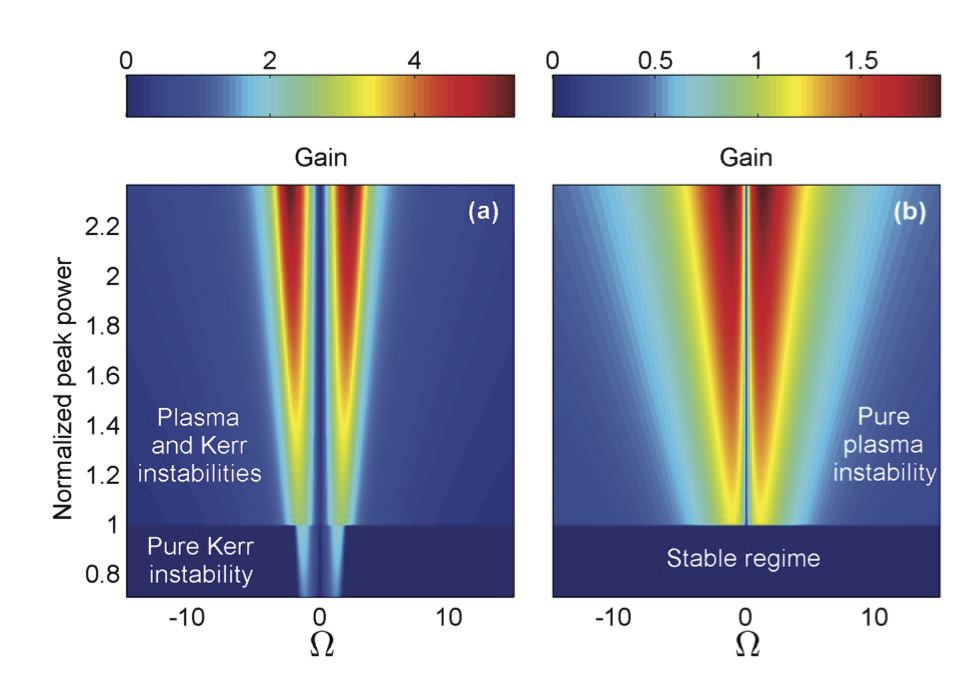}
\caption{(Color online). MI spectral gain profile versus normalized input peak power in (a) anomalous and (b) normal dispersion regimes with equal magnitude of GVD. For the anomalous regime, a CW with $\lambda=1064$ nm is launched into a Kagome Ar-filled HC-PCF with core diameter $20$ $\mu$m, gas pressure of 1 bar, and $\beta_{2} \simeq-2.8$ ps$ ^{2} $/km. In this fiber, threshold ionization power is $ \simeq105$~MW at room temperature. For the normal regime simulation, we assume hypothetically a gas-filled fiber with a $\beta_{2}$ of the same magnitude as in (a), but with a positive sign.
\label{Fig4-7}}
\end{figure}

\paragraph*{Plasma-induced modulational instability---}  After the initial SPM stage is over, the interplay between nonlinear and dispersive effects can lead to an instability that modulates the temporal profile of the pulse, creating new spectral sidebands referred to as modulational instability (MI) \cite{Hasegawa80,Tai86a,Tai86b}. Starting from Eq. (\ref{nGNLSE2}), MI due to the photoionization nonlinearity can be investigated by using the standard approach  described in \cite{Agrawal07,Hasegawa80}. The Kerr-induced MI occurs only in the anomalous dispersion regime over a defined bandwidth \cite{Agrawal07}. However, the presence of the photoionization process induces an unusual instability that can exist {\em in both normal and anomalous dispersion regimes, and for any frequency}. The spectral dependence of the gain of these instabilities on different peak powers is shown in Fig. \ref{Fig4-7} for (a) anomalous and (b) normal dispersion regimes, where the physical powers are normalized to the threshold ionization power, i.e., $ |\psi|^{2} _{\mathrm{th}} =1 $. When the normalized input power $ \psi_{0}^{2}\leq |\psi|^{2}_{\mathrm{th}} $, we have the traditional side-lobes, which exist uniquely in the anomalous dispersion regime, due to the Kerr-nonlinearity. However when $ \psi_{0}^{2} > |\psi|^{2}_{\mathrm{th}} $, photoionization-induced instability generates unbounded side-lobes with slowly-decaying tails. A similar situation occurs in the normal regime, however, the gain is slightly lower due to the absence of the Kerr contribution. For this case there are no instabilities below the threshold power since no plasma is generated.

\begin{figure}
\includegraphics[width=8.6cm]{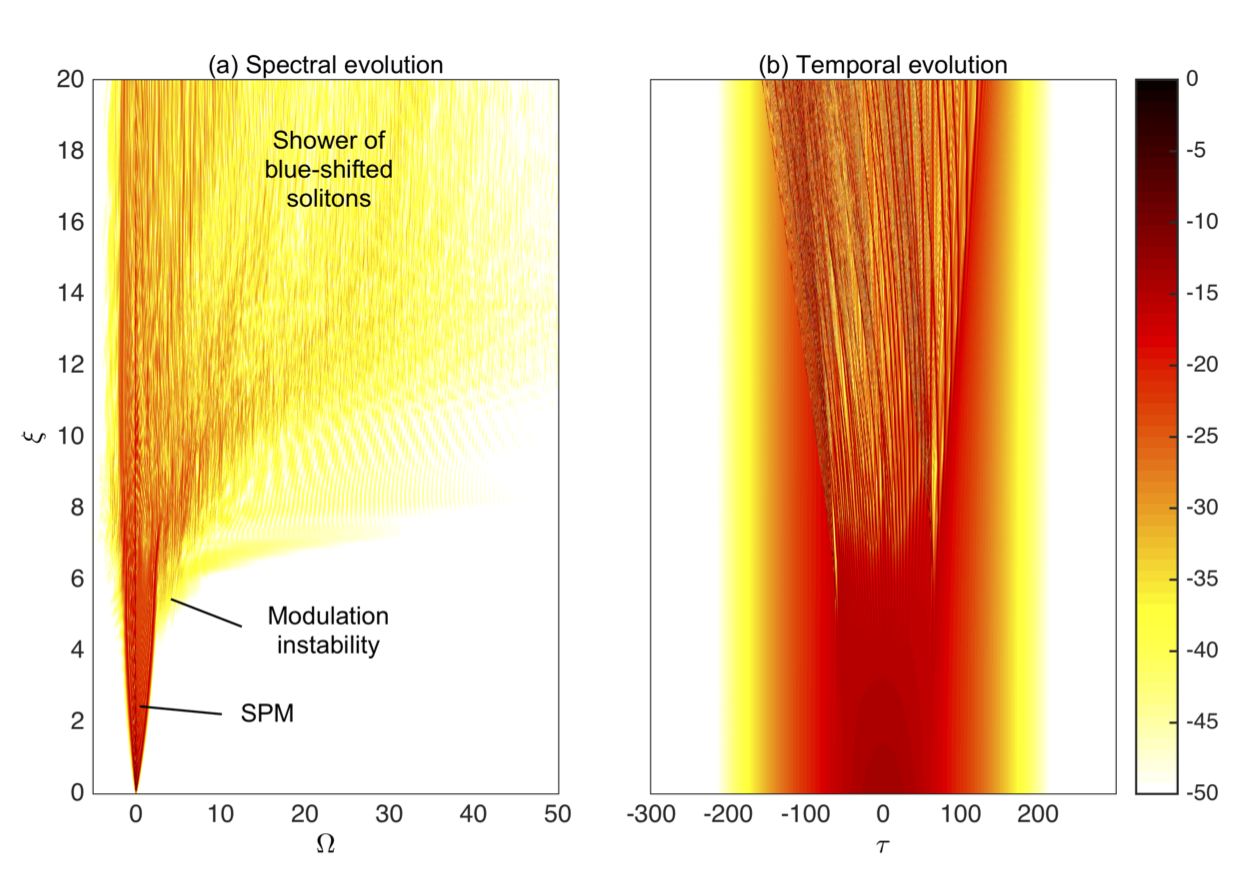}
\caption{(a) Spectral and (b) temporal evolution of a long Gaussian pulse propagating in an Ar-filled HC-PCF with a gas pressure 1 bar, and a hexagonal core-diameter 20 $ \mu $m. The input pulse has a peak power 200 MW and a duration 0.707 ps 1/e-intensity half-width. Contour plots are given in a logarithmic scale and truncated at -50 dB.\label{Fig4-8}}
\end{figure}

\begin{figure}
\includegraphics[width=8.6cm]{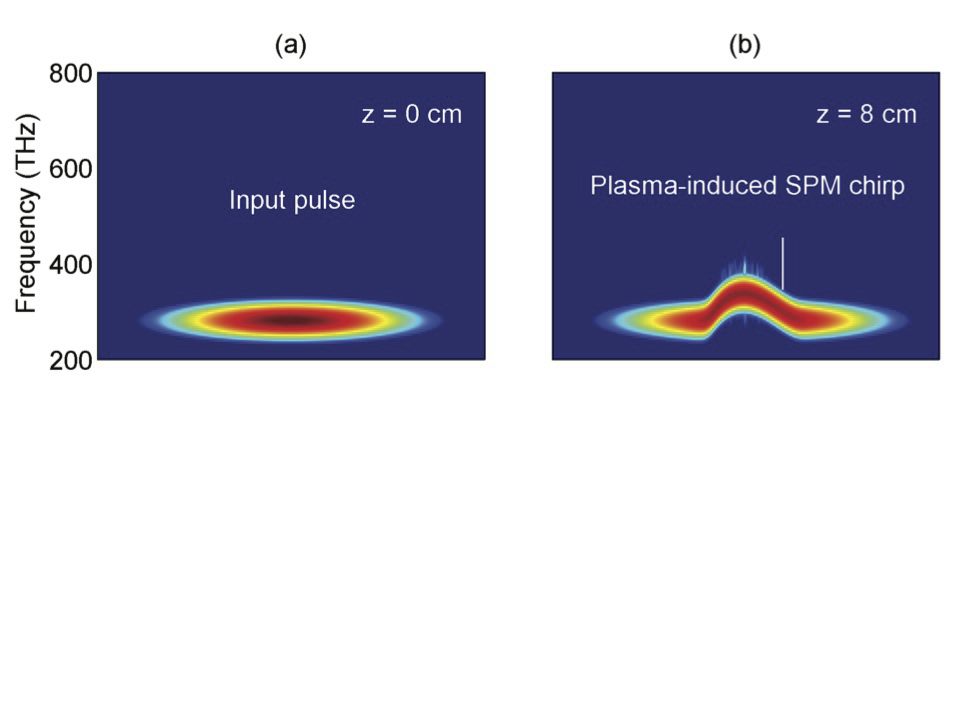}\vspace{-3cm}
\includegraphics[width=8.6cm]{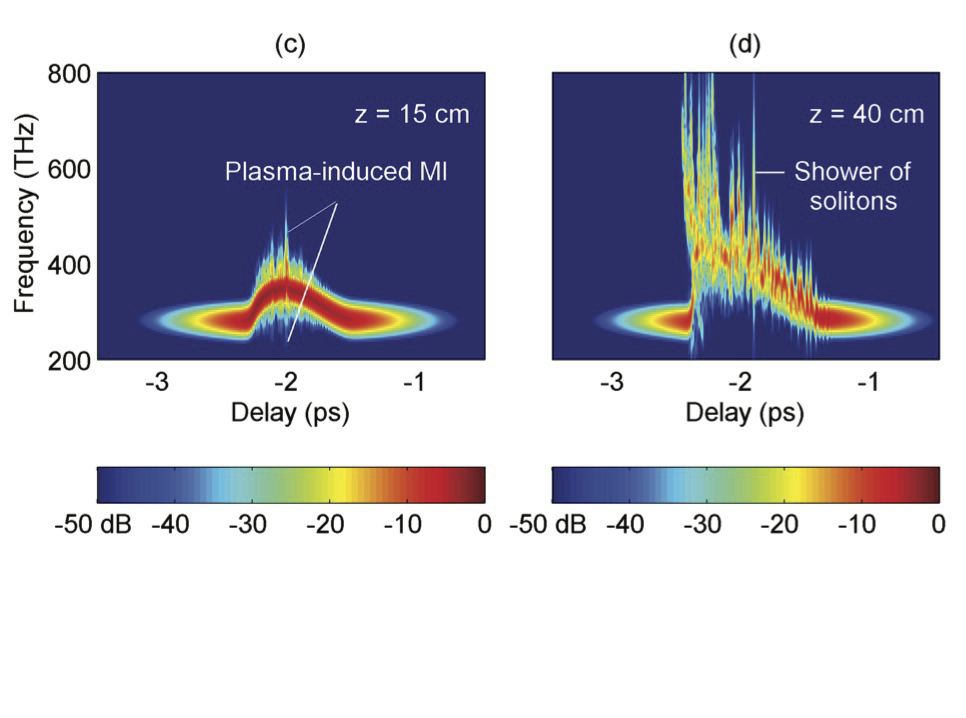}\vspace{-1.7cm}
\caption{(Color online). XFROG spectrograms for the propagation through an Ar-filled HC-PCF. The simulation parameters are the same as in Fig. \ref{Fig4-8}. The reference pulse is a Gaussian with FWHM 50 fs. (a) Input pulse. (b) SPM frequency chirping. (c) Modulational instability. (d) Pulse disintegration into multiple blueshifting solitons.
\label{Fig4-9}}
\end{figure}

Propagation of a long Gaussian pulse inside an anomalous dispersive gas-filled HC PCF is portrayed in the panels (a,b) of Fig. \ref{Fig4-8}, obtained by simulating Eq. (\ref{nGNLSE2}). The first stage of propagation shows asymmetric spectral broadening towards the blue due to ionization-induced SPM. Immediately after the SPM stage, dispersion starts to play a role, and due to the combined Kerr and ionization MIs, broad and slowly decaying side lobes are generated and amplified quickly. In the third and final stage in the propagation, strongly blueshifted solitons are emitted. Further insight into the dynamics can be obtained from Fig. \ref{Fig4-9}, where the evolution of the XFROG spectrograms of the pulse at different positions along the fiber is shown. The pulse is initially asymmetrically chirped in the center of the pulse towards high frequencies, [Fig. \ref{Fig4-9}(b)], due to the higher plasma density created at the peak intensities.  At the same time two imbalanced ionization-induced MI sidebands appear in the pulse spectrum [Fig. \ref{Fig4-9}(c)]. MI facilitates the formation of many solitons. In less than half a meter of propagation the initial pulse disintegrates into a `shower' of solitary waves, see Fig. \ref{Fig4-9}(d), each undergoing a strong self-frequency blueshift induced by the intrapulse photoionization.

\section{Conclusions and final remarks}
In this review, we have presented very recent results concerning soliton dynamics in gas-filled HC-PCFs. We have divided gases into two main categories: Raman-active (molecular gases) and Raman-inactive (monoatomic gases). The former kind of gases, such as molecular hydrogen, are characterized by long molecular coherence in comparison to silica glass. This results in highly non-instantaneous interactions that can be detected by launching a probe pulse, delayed from the main pump pulse. For a weak probe, the problem is reduced to the motion of a quantum particle in a periodic `temporal' crystal subject to an external force. Phenomena related to condensed matter physics such as Wannier-stark ladder, Bloch oscillations, and Zener tunneling are predicted to occur. However, if the probe is another ultrashort intense soliton, phenomena such as soliton oscillations and transport have been shown to occur. Moreover, in this case the temporal crystal is upgraded to a spatiotemporal one with a uniform or chirped spatial period. In Raman-inactive gases, such as argon, we have investigated the effect of photoionization effects on the evolution of short and long pulses. Unique phenomena such as soliton self-frequency blueshift, asymmetrical self-phase modulation, universal modulation instability, and shower of blueshifted solitons have been studied and elaborated. These fresh theoretical results, supported by recent experiments, can  pave the way for the manipulation and control of the pulse dynamics in PCFs for demonstrating  completely novel optical devices.

M. Saleh would like to acknowledge the support of Royal Society of Edinburgh and Scottish Government.


%

\end{document}